\overfullrule 0pt
\vsize=9truein
\hsize=6.5truein

\def \A {{\hbar \omega^3 \over 2\pi^2 c^3}}
\def \Ap {{\hbar \omega'^3 \over 2\pi^2 c^3}}
\def \Is {$I_*$}
\def \It	{$I_{\tau}$}
\def \Idt	{$I_{\Delta\tau}$}
\def \FF {{\bf F}}
\def \F	{{\bf f}}
\def \E	{{\bf E}}
\def \B	{{\bf B}}
\def \N	{{\bf N}}
\def \a	{{\bf a}}
\def \p  {{\bf p}}
\def \g  {{\bf g}}

\def \v  {{\bf v}}
\def \w {\omega}
\def \l {\lambda}
\def \ew {\eta(\omega)}
\def \ewp {\eta(\omega')}
\def \Nt	{{\bf N}_{\tau}}

\def \Nzp	{{\bf N}^{zp}}
\def \Fzp	{{\bf f}^{zp}}
\def \Ezp	{{\bf E}^{zp}}
\def \Bzp	{{\bf B}^{zp}}
\def \pzp	{{\bf p}^{zp}}
\def \gzp	{{\bf g}^{zp}}
\def \D {\Delta}
\def \csh {\cosh \left( {a\t \over c} \right)}
\def \cxh {\cosh^2 \left( {a\t \over c} \right)}
\def \tnh {\tanh \left( {a\t \over c} \right)}
\def \txh {\tanh^2 \left( {a\t \over c} \right)}
\def \snh {\sinh \left( {a\t \over c} \right)}
\def \snx {\sinh \left( {2 a\t \over c} \right)}
\def \csA {\cosh(A)}
\def \cxA {\cosh^2(A)}
\def \snA {\sinh(A)}
\def \tnA {\tanh(A)}
\def \txA {\tanh^2(A)}
\def \t {\tau}
\def \b {\beta}
\def \bt {\beta_{\tau}}
\def \gt {\gamma_{\tau}}
\def \< {\left\langle}
\def \> {\right\rangle}
\def \[ {\left[ }
\def \] {\right] }
\def \( {\left(}
\def \) {\right)}
\def \lb {\left\{}
\def \rb {\right\}}
\def \Sm { \sum_{\lambda=1}^2 }
\def \Smp { \sum_{\lambda'=1}^2 }
\def \Ik { \int d^3k \ }
\def \Ikp { \int d^3k' \ }
\def \Ikk { \int {d^3k' \over k'} }
\def \he {\hat{\epsilon}}
\def \hk {\hat{k}}
\def \hep {\hat{\epsilon}'}
\def \hkp {\hat{k}'}
\def \ke { (\hat{k} \times \hat{\epsilon}) }
\def \kep { (\hat{k}' \times \hat{\epsilon}') }
\def \ra {\rightarrow}
\def \Hw {H_{zp}^2(\omega)}
\def \Hwp {H_{zp}^2(\omega')}
\def \hf {{1 \over 2}}
\def \sn {\smallskip\noindent}

\centerline{\bf Contribution to inertial mass by reaction of the vacuum}
\centerline{\bf to accelerated motion }
\bigskip
\centerline{Alfonso Rueda}
\centerline{\it Department of Electrical Engineering, ECS Building}
\centerline{\it California State University, 1250 Bellflower Blvd.,
Long Beach, California 90840}
\centerline{arueda@csulb.edu}
\bigskip
\centerline{Bernhard Haisch}
\centerline{\it Solar and Astrophysics Laboratory, Dept. H1-12, Bldg.
252,  Lockheed Martin}
\centerline{\it 3251 Hanover Street, Palo Alto, California 94304}
\centerline{and}
\centerline{\it Max-Planck-Institut f\"ur Extraterrestrische Physik,
D-85740 Garching, Germany}
\centerline{haisch@starspot.com}
\bigskip
\centerline{($\copyright$ 1998 Foundations of Physics, to appear in 3rd quarter)}
\bigskip
\centerline{\bf Abstract}
\bigskip We present an approach to understanding the origin of inertia involving
the electromagnetic component of the quantum vacuum and propose this as a step
toward an alternative to Mach's principle. Preliminary analysis of the momentum flux
of the classical electromagnetic  zero-point radiation impinging on accelerated
objects as viewed by an inertial observer suggests that the resistance to
acceleration attributed to inertia may be at least in part a force of opposition
originating in the vacuum.  This analysis avoids the {\it ad hoc} modeling of
particle-field interaction dynamics used previously by Haisch, Rueda and Puthoff
(Phys. Rev. A {\bf 49}, 678, 1994) to derive a similar result.  This present
approach is not dependent upon what happens at the particle point, but on how an
external observer assesses the kinematical characteristics of the zero-point
radiation impinging on the accelerated object.  A relativistic form of the equation
of motion results from the present analysis.  Its manifestly covariant form yields
a simple result that may be interpreted as a contribution to inertial mass. We note
that our approach is related by the principle of equivalence to Sakharov's
conjecture (Sov. Phys. Dokl. {\bf 12}, 1040, 1968) of a connection between Einstein
action and the vacuum.  The argument presented may thus be construed as a
descendant of Sakharov's conjecture by which we attempt to attribute a mass-giving
property to the electromagnetic component --- and possibly other components --- of
the  vacuum. In this view the physical momentum of an object is related to the
radiative momentum flux of the vacuum instantaneously contained in the
characteristic proper volume of the object.  The interaction process between the
accelerated object and the vacuum (akin to absorption or scattering of
electromagnetic radiation) appears to generate a physical resistance (reaction
force) to acceleration suggestive of what has been historically known as inertia.
\bigskip\noindent PACS:   03-65-W;    03.20 + I;   03.50-K;   95.30 Sf

\vfill\eject
\centerline{\bf I. INTRODUCTION}
\bigskip As discussed recently by Vigier [1], ``the origin and nature of inertial
forces\dots can be considered as an unsolved mystery in modern physics. It still
sits, like Banco's ghost, at any banquet of natural philosophers.''
The instantaneous opposition to acceleration of all material
objects is conventionally assumed to be a universal property of
matter known as inertia.  Historically there have been two views on the
origin of inertial mass.  It has been assumed to be either an
inherent internal property of matter capable of no further
explanation, or, in the view of Mach, a property that somehow
originates externally in a collective linkage among all matter in the
universe.  This last, often referred to as Mach's principle, may be
exemplified in a thought experiment.  Rotation is a form of
acceleration.  The inertia of matter manifests itself in the
existence of centrifugal (and Coriolis) forces in the reference frame
of a rotating object.  Imagine a universe containing only a single
object.  In the view of Mach it would be an absurdity to claim that,
in an otherwise empty universe, this object is capable of rotation. 
This would imply that centrifugal (and Coriolis) forces could not
manifest and that therefore the single object in an empty universe
should be devoid of inertia.  If a single external object is now
introduced, the phenomenon of rotation, by virtue of external
reference, is again possible and the inertia of the rotating object
should reappear.  This allows the interpretation that the external
object is the cause of  the inertia of the rotating object.  However
it would be unphysical to assume that any external object no matter
how minute should be capable of creating all at once the ``full
inertia'' in the rotating object that it would otherwise possess in
the ``standard'' universe.  It can thus be argued in the Machian view
that inertia must be an asymptotic function of surrounding matter
that would gradually come into being as the universe is filled around
the object in question.

\bigskip A rigorous and quantitative formulation of Mach's principle
has never been successfully developed [2].  A tentative attempt by
Sciama [3] to quantify Mach's principle by associating inertial mass
generation with a vector extension of the gravitational potential
(analogous to a gravitation current) resulted in a prediction that
was later shown to be inconsistent with experimental evidence.  In
the Sciama formulation, the asymmetrical distribution of surrounding
matter in the Milky Way should result in a directional dependence
of inertial mass with respect to galactic coordinates as measurable
in a laboratory, this amounting to a variation on the order of
$\Delta m/m = 10^{-7}$ whereas the experiments of Hughes and Drever
subsequently indicated that $\Delta m/m \le 10^{-20}$ [4].

\bigskip That general relativity is not Machian is exemplified by the
fact that it is possible to formulate solutions of the field
equations for an empty universe and for a rotating universe.  Recent examination on the
relationship of the Lense-Thirring effect with general relativity further
demonstrates the absence of a clear relationship and possible inconsistency between
Mach's principle and general relativity [5]. Additional conflicts between general
relativity and Mach's principle are presented by Vigier [1].

\bigskip The Machian view would
imply that it is entirely arbitrary whether one regards acceleration as motion (or
rotation) of the object in question or as counter-motion (or
counter-rotation) of the rest of the universe.  However because the
inertia reaction force occurs at the same instant that acceleration
is applied to an object it becomes causally awkward to explain how
Mach's principle could be accommodated without the need for
instantaneous propagation of some kind of back-reaction field involving infinite
velocity, thus violating relativity and causality [6]. Preservation of causality
is, of course, a strong argument for finding a basis of inertia that involves
locally-originating forces and interactions. The approach of Vigier [1] is to find
such a non-Machian basis in local interactions of a real Dirac subquantum aether
model stemming from Einstein-de Broglie-Bohm causal stochastic quantum mechanics.
The view presented herein substitutes for Mach's principle in identifying the
electromagnetic fields of the quantum vacuum as the external causative agent of
inertia by providing a locally-originating reaction force. 
Limitations of treatment allow us to show this only for the
electromagnetic vacuum, leaving the contributions of other vacuum
fields for further extensions of the theory. An example of a contribution by other
vacuum fields is precisely the one recently presented by Vigier [1].

\bigskip The original development of this idea [7] proposed that the inertial
property of matter could originate in Lorentz-force interactions between
electromagnetically interacting particles at the level of their most fundamental
components (e.g., electrons and quarks) and the quantum vacuum (QV) [8].
This general idea is a descendent of a conjecture of Sakharov for the
case of Einstein action [9] that can be extended by the principle of
equivalence to the case of inertia.  The approach of stochastic
electrodynamics (SED) was used in [7] to study the classical dynamics
of a highly idealized model of a fundamental particle constituent of
matter (that contained  a ``parton'', i.e., a surrogate for a very
fundamental particle component) responding to the driving forces of
the so-called electromagnetic zero-point field (ZPF), the classical
analog of the QV.

\bigskip Reservations can be raised about the proposal of  [7], for
example:  (a) the complexity, in that the analysis of the classical
charged particle-ZPF dynamics involved an extensive calculational
development which complicated the assessment of the physical validity
of the approach; and (b) the introduction of {\it ad hoc} dynamical
models for the interaction of the field and matter particularly at
very high frequencies.  Among these dynamical assumptions there were
two of clear concern:  (i) the idealized representation of a particle
as a ``parton'' (Planck oscillator) and (ii) the use of the
Abraham-Lorentz-Dirac (ALD) equation as the starting point (the ALD
equation originates in Newton's Laws).  Details on problems with the
development in [7] will be discussed in [10].

\bigskip Consequently the primary purpose of the present paper is to
outline a simple approach which  avoids some of these
model-related issues by examining how an opposing flux of radiative
energy and momentum should arise under natural and suitable
assumptions  in an accelerated frame from the viewpoint of an
inertial observer without regard to details of particle-field
dynamics, i.e., independently of any dynamical models for particles. As the details
of particle-field interactions are not of concern in the present case, the use of
the classical electromagnetic ZPF formalism of SED looks quite natural. Using
standard relativistic transformations for the electromagnetic fields, it is argued
that upon acceleration a time rate of change of momentum density or momentum flux
will arise out of the ZPF in the proper volume of any accelerating object, and that
this turns out to be directed against and linearly proportional to the
acceleration  (Sec. II, III, IV and V).  This arises after evaluation of the ZPF
momentum density (Sec. IV, V) as it appears at a given point in an accelerated
frame $S$, to an independent inertial laboratory observer due to transformations of
the fields from the observer's inertial laboratory reference frame \Is, to another
inertial frame \It, instantaneously comoving with the object (Sec. II) and from the
viewpoint of the observer in the laboratory inertial frame \Is. 
Absorption or scattering of this radiation by the accelerated charged
particle is found to result in a force opposing the acceleration,
yielding an  $\F=m\a$ relation for subrelativistic motions.  We
follow standard notation in using $\F$ to refer to a three-force, and
$\FF=\gamma \F$ to refer to the corresponding space part of the
relativistic four-force; cf. Eq. (9). The relativistic form of the
force expression is obtained in Appendix D and presented in Sec. VI. 
For the execution of this development we again assume hyperbolic
motion [7, 11, 12] (i.e., constant proper acceleration).  Extension
to an arbitrary accelerated motion is readily envisioned (Sec. VII). 
Section VIII concludes the paper.  Important details or elaborations,
extensions, and refinements are left for the appendices (A, B, C and
D).

\bigskip What we tentatively propose here is that when an object of
rest mass $m$ receives an impulse and is thereby accelerated by an
external agent, the following two features deserve special attention.

\bigskip (1) The scattering of the incoming ZPF flux within the object is what
generates a reaction force heretofore attributed solely to the existence of an
unexplained  property called inertia. This clearly must be directed opposite to the
direction of acceleration. As shown below this reaction force will turn out to be
proportional to the acceleration $\a$ but in opposite direction to it, $\Fzp \sim
-\a$.  If there were an ideal body acting as a ``perfect absorber'',  i.e,
capable of interacting with all the incoming flux of momentum from
the ZPF up to the highest frequencies, an enormous maximum total
reaction force would appear.  In the case of a more realistic but
still idealized body of characteristic proper volume $V_0$ that
intercepts only a certain proportion $\eta$ of the radiation ($0 <
\eta< 1$), owing to this  opacity  there appears a reaction force on
the body  of the form  $\Fzp \sim - V_0 \eta \a$.  The effect should
clearly be larger for bodies of larger volume $V_0$ and/or such that 
$\eta$, the matter-radiation coupling coefficient, is larger.  In
this interpretation inertial mass becomes a function of such
opacity.  As the SED classical ZPF background is such a large
reservoir of energy, this phenomenological coupling coefficient
$\eta$ can be extremely small for a substantial effect to still
appear.  We do not concern ourselves in this paper with the nature or
strength of $\eta$,  i.e., we omit consideration of the detailed
dynamics of interaction of the ZPF with matter in general or with
material particles in particular.  We report only on the necessary existence of
a force of opposition by the ZPF as characterized by a change in the
electromagnetic momentum density to the accelerated motion of the
object without any concern for the details of the particle-radiation
interaction embodied in the efficiency factor $\eta$.

\bigskip (2) After the acceleration process is completed, from the
point of view of an inertial observer attached to the stationary
laboratory frame there appears associated with the body in motion a
net flux of momentum density in the surrounding ZPF. In other words,
on calculating the ZPF momentum contained in the object as referenced
to the observer's own inertial frame, the observer would conclude
that a certain amount of momentum is instantaneously contained within
the proper volume, $V_0$, of the moving object. This momentum is
directly related to what would normally be called the physical
momentum of the object. Calculated with respect to its own frame the
object itself would find no net ZPF momentum contained within itself,
consistent with the view that one's own momentum is necessarily
always zero.

\vfill\eject
\centerline{\bf II. ZERO-POINT FIELD AND HYPERBOLIC MOTION}

\bigskip In the following we reference a small (in a sense to be
specified below) and accelerated ``object'' consisting of elementary
``particles'' contained within a small volume. The term ``object''
can refer to either such a spatially extended but small entity or, in
the context of reference frames, to its central point. We assume a
non-inertial frame of reference, $S$, accelerated in such a way that
the acceleration $\a$ as seen from an object fixed to a specific
point, namely $(c^2/a, 0, 0)$, in the accelerated system, $S$,
remains constant, i.e., the point $(c^2/a, 0, 0)$ is uniformly
accelerated.  Such condition leads as in [7, 11] to the well-known
case of hyperbolic motion [12].  We again represent the classical
electromagnetic ZPF in the traditional form and assume the same three
reference systems \Is, \It, and $S$ as in [7] and originally
introduced in [11].  \Is \ is the inertial laboratory frame.  $S$ is
the accelerated frame in which the object is placed at rest at the
point $(c^2/a, 0, 0)$.  $\t$ is the object proper time as measured by
a clock located at this same object point $(c^2/a, 0, 0)$ of $S$. 
\It \ is an inertial system whose $(c^2/a, 0, 0)$ point at proper
time $\t$ exactly (but only instantaneously) coincides with the
object point of $S$.  The acceleration of this $(c^2/a, 0, 0)$ point
of $S$ is $\a$ as measured from \It.  Hyperbolic motion is defined
such that $\a$ is the same for all proper times $\t$ as measured in
the corresponding \It \  frames at a point $(c^2/a, 0, 0)$ that in
each one of these \It \ frames instantaneously comoves and coincides
with the corresponding object point, namely $(c^2/a, 0, 0)$ of $S$. 
At proper time $\t= 0$ this object point  of the $S$-system  also
instantaneously coincides with the $(c^2/a, 0, 0)$ point of \Is \ and
thus \Is = \It$(\t = 0)$. We refer to the observer's laboratory time
in \Is \ as $t_*$, chosen such that $t_*=0$ at $\t = 0$.  For
simplicity we let the object acceleration $\a$ at proper time $\t$
take place along the $x$-direction so that  $\a= a\hat{x}$ is the
same constant vector, as seen at every proper time $\t$ in every
corresponding \It \ system.  The acceleration of the $(c^2/a, 0, 0)$
point of $S$ as seen from \Is \ is $\a_*=\gamma_{\t}^{-3}\a$ [12]. 
Occasionally we refer to $S$ as the Rindler non-inertial frame.  We
take it as a ``rigid'' frame [12].  It can be shown that as a
consequence the acceleration $\a$ is not the same for the different
points of $S$, but we are only interested in points inside a {\it
small} neighborhood of the center of the accelerated object [12]. Specifically we
are interested in a neighborhood of the object's central point that
contains the object and within which the acceleration is everywhere
essentially the same.

\bigskip Because of the hyperbolic motion [7,11,12], the velocity
$u_x(\t) = \bt c$ of the object point fixed in $S$ with respect to
\Is,  is

$$ \bt={u_x(\t) \over c} = \tnh  
\eqno(1)$$

\sn and then

$$ \gamma_{\t} = \left( 1-\beta_{\t}^2 \right)^{-1/2} = \csh .  
\eqno(2)$$

The ZPF in the laboratory system \Is \ is given by [7,11]

$$\Ezp ({\bf R}_*,t_*) = \Sm \int d^3 k \
\he({\bf k}, \l) H_{zp}(\omega) \cos [ {\bf k \cdot R}_* - \omega t_*
- \theta({\bf k},\l) ] ,
\eqno(3a)$$

$$\Bzp ({\bf R}_*,t_*) = \Sm \int d^3 k \ (\hk \times \he)
H_{zp}(\omega) \cos [ {\bf k \cdot R}_* - \omega t_* - \theta({\bf
k},\l) ] .
\eqno(3b)$$

\sn (See however statements on a normalization factor following Eq.
(A5) in Appendix A. See also Ref. [13]).  ${\bf R}_*$ and $t_*$ refer
respectively to the space and time coordinates of the point of
observation of the  field  in  \Is.   At  $t _* = 0$, the point 
${\bf R}_* = (c^2/a)\hat{x}$ of \Is \ and the object in $S$ coincide 
(see Eq. (6) below).   The phase term $\theta({\bf k},\l)$ is a
family of random variables, uniformly distributed between 0 and
2$\pi$, whose mutually independent elements are indexed by the
wavevector  ${\bf k}$ and the polarization index $\l$ (or more
technically, $\theta({\bf k},\l)$ is a stochastic process with index
set $\{ ({\bf k},\l) \}$). Furthermore one defines,

$$H_{zp}^2(\omega)={\hbar \omega \over 2 \pi^2} . 
\eqno(4)$$

\sn As the coordinates ${\bf R}_*$ and time $t_*$ refer to the
particle point of the accelerated frame $S$ as viewed from \Is \ we,
for convenience, Lorentz-transform the fields from \Is \ to the
corresponding \It \  frame tangential to $S$ and then (omitting for
simplicity to display explicitly the $\l$ and ${\bf k}$ dependence in
the polarization vectors, $\he = \he({\bf k}, \l)$), obtain

$$
\Ezp (0,\t) = \Sm \int d^3 k \
\big\{
\hat{x}\he_x +
\hat{y}\gamma_{\t}  [\he_y - \beta_{\t} (\hk\times \he)_z]+
\hat{z}\gamma_{\t}  [\he_z + \beta_{\t} (\hk\times \he)_y]
\big\}
$$
$$
\times H_{zp}(\omega) \cos [ {\bf k \cdot R}_* - \omega t_* -
\theta({\bf k},\l) ]
\eqno(5a)$$

$$
\Bzp (0,\t) = \Sm \int d^3 k \
\big\{
\hat{x}(\hk \times \he)_x +
\hat{y}\gamma_{\t}  [ (\hk\times \he)_y + \beta_{\t}\he_z]+
\hat{z}\gamma_{\t}  [ (\hk\times \he)_z - \beta_{\t}\he_y]
\big\}
$$
$$
\times H_{zp}(\omega) \cos [ {\bf k \cdot R}_* - \omega t_* -
\theta({\bf k},\l) ] ,
\eqno(5b)$$

\sn where the zero in the argument of the \It \ fields,
$\Ezp$ and $\Bzp$ actually means the \It \  spatial point  $(c^2/a,
0, 0)$.  Here we observe three things.  First, we take the fields
that correspond to the ZPF as viewed from every inertial frame \It \ 
(whose $(c^2/a, 0, 0)$ point coincides with the particle point
$(c^2/a, 0, 0)$ of $S$ and instantaneously comoves with the object at
the corresponding instant of proper time $\t$) to also represent the
ZPF viewed instantaneously and from the single point $(c^2/a, 0, 0)$
in $S$.  Note the dependence on the proper time $\t$ of the object. 
Second, the corresponding fields in \It \ are obtained from a simple
Lorentz rotation from \Is \ into \It.   Hence for every proper time
$\t$ in the Rindler non-inertial frame, the fields
$\Ezp$ and $\Bzp$ appear as expanded in terms of the four-vector
$(k,{\bf k})$ whose components are the wavevector magnitude $k =
\omega/c$ and wavevector ${\bf k}$ of \Is \ [11].  This will prove to
be an advantageous simplifying feature that will help in establishing
\Is \ as the ultimate reference frame in terms of which everything at
the object point in $S$ at any proper time $\t$ is written.

\bigskip The third and final point is crucial:  Though the fields at the object
point in $S$ and in the corresponding tangential frame \It \
instantaneously coincide,  this does not mean that detectors in $S$
and in \It \ will be subject to the same effect, i.e., experience the
same radiation-field time evolution.  {\it Detectors need time to
perform their measurements}: This necessarily involves integration
over some interval of time and the evolution of the fields in $S$ and
in \It \ are obviously different.  Hence a detector at rest in \It \
and  the same detector at rest in $S$ do not experience the same
thing even during a short time interval.  (This point touches on the
origin of the Unruh-Davies effect, which is beyond the scope of the
present paper.) 

\bigskip Consider any dynamical system such as, for example, a
collection of interacting particles:  An infinite number of different
dynamical states can yield the same instantaneous snapshot of the
system configuration even though the dynamical states of motion may
be radically different.  Two snapshots separated in time are
necessary to distinguish differences among systems coordinates
involving velocities; three will begin to distinguish accelerations,
etc.   The state of the system cannot be captured in an exact instant
of time, but rather is a function of the time evolution. 
Summarizing, while the two fields, namely that of $S$ and that of
\It, are the same at a given space-time point, the evolution of the
field in $S$ and the evolution of the field in \It \  are by no means
the same.  Furthermore any field or radiation measurements in \It \
and in $S$ both take some time and are not confined to a single
space-time point.

\bigskip We clarify the notation used in the sense that all
polarization components are understood to be scalars, i.e.,
directional cosines, but written in the form
$\he_i ({\bf k},\l) \equiv \he \cdot \hat{x}_i$, where 
$\hat{x}_i=\hat{x},\hat{y},\hat{z}; \ i=x,y,z$. The karat means that
they come from axial projections of the polarization unit vector
$\he$ .  We use the same convention  for components of the $\hk$ unit
vector where, e.g., $\hk_x$ denotes $\hk \cdot \hat{x}$.  We can
select space and time coordinates and orientation in \Is \ such that
[11,12]

$${\bf R}_*(\t)\cdot \hat{x} = {c^2 \over a} \csh 
\eqno(6)$$

$$t_*={c \over a} \snh
\eqno(7)$$

From Eqs. (1), (2) and (6--9) one obtains [7, 11]

$$
\Ezp (0,\t) = \Sm \int d^3 k \times
$$
$$
\bigg\{
\hat{x}\he_x +
\hat{y}\csh  \[ \he_y - \tnh (\hk\times \he)_z \] +
\hat{z}\csh  \[ \he_z + \tnh (\hk\times \he)_y \]
\bigg\}
$$
$$
\times H_{zp}(\omega) \cos \[ k_x{c^2 \over a} \csh - {\w c \over a}
\ \snh - \theta({\bf k},\l) \]
\eqno(8a)$$

$$
\Bzp (0,\t) = \Sm \int d^3 k \times
$$
$$
\bigg\{
\hat{x}(\hk \times \he)_x +
\hat{y}\csh  \[ (\hk\times \he)_y + \tnh \he_z \] +
\hat{z}\csh  \[ (\hk\times \he)_z - \tnh \he_y \]
\bigg\}
$$
$$
\times H_{zp}(\omega) \cos \[ k_x{c^2 \over a} \csh - {\w c \over a}
\ \snh - \theta({\bf k},\l) \] .
\eqno(8b)$$

\sn This is the ZPF as instantaneously viewed from the object fixed
to the point $(c^2/a, 0, 0)$ of $S$ that is performing the hyperbolic
motion.

\vfill\eject
\centerline{\bf III. DYNAMICAL SETTING: FORCE OF RESISTANCE TO}
\centerline{\bf ACCELERATED MOTION AND ZPF MOMENTUM FLUX}

\bigskip In 1968 Sakharov [9] published a brief conjecture regarding Einstein
action. In this approach the concept of mass does not arise; objects simply move
along geodesics. However one could go on to interpret this as
gravity arising in certain perturbations by massive bodies of the surrounding vacuum
fields, and the principle of equivalence would then imply that this should extend to
inertia.  A successful extension to a more general field theory than
the one contemplated here would likely suggest that an inertia effect
is caused by the vacuum  of that theory, without the need of
postulating any additional field solely for the purpose of giving
mass to material entities (e.g. a mass-giving Higgs-type field).

\bigskip The objective of the present paper is to study, by a
completely different approach to that in [7],  the  hypothesis that
inertial mass may be considered a vacuum effect, i.e., that within
the limited context of our treatment, the parameter $m$ of Newton's
second law $(\F=m\a)$ can be explained as an effect due to the ZPF,
i.e., that the inertial rest mass can be explained, at least in part, as a
coefficient involving the object, its electromagnetic coupling and other ZPF
parameters.  This is  attempted now not by means of a dynamical
analysis on a very specific model (as in [7]) but instead by careful
examination of the structure of the fields viewed in relation to an
object being compelled to perform accelerated motion by an external
agent.  The goal of such analysis is to find an expression and an
explanation for the $m$ parameter.

Newton's second law can be more
generally written, but still in its traditional nonrelativistic form,
as

$$\F={d\p \over dt}=\lim_{\D t \ra 0} {\D \p \over \D t}  ,
\eqno(9a)$$

\sn which is the limiting form of the space part of the relativistic
four-force form of Newton's law:

$$\FF={d\p \over d\t}= \gamma{d\p \over dt} ,
\eqno(9b)$$

\sn which for the case when $\beta \ra 0$ and $\gamma \ra 1$
(corresponding for us to the object in the $\t \ra 0$ limit when \It
\ coincides with \Is \ and then $\gt \ra 1$) becomes

$$\F={d\p \over d\t}\bigg| _{\t=0} .
\eqno(9c)$$

\bigskip Having defined force in his second law as the rate of change
of momentum imparted to an object by an agent, Newton then states in
his third law that such a force will result in the creation of an
equal and opposite reaction force back upon the accelerating agent. 
The concept of inertia becomes then a necessity: Inertia is thus
necessarily attributed to the accelerating object in order to
generate the equal and opposite reaction force upon the agent
required by the third law.  It is our proposition that resistance
from the vacuum is what physically provides that reaction force.  One
can interpret this as either the origin of inertia of matter or as a
substitute for the concept of innate inertia of matter.  In other
words, inertia becomes in a sense a placeholder for this heretofore
undiscovered vacuum-based reaction force which is a necessary
requirement of Newton's third law.  Force is then seen to be a
primary concept; inertia is not.

\bigskip Newton's third law is essentially a statement about symmetry
in nature for contact forces. In the static case (e.g., pressing one hand against
the other), from symmetry alone an applied force $\F$, must necessarily result in a
reaction force $\F_r$ such that

$$\F=-\F_r .
\eqno(10)$$

\sn Inertia as the dynamical extension of this law can be made
explicit by writing the $\F=m\a$ relation as

$$\F=-(-m\a) ,
\eqno(11)$$

\sn which makes it clear that inertia as a resistance to acceleration
is equivalent to a reaction force of the form

$$\F_r=-m\a .
\eqno(12)$$

\bigskip To recapitulate our argument, Newton's third law states that
if an agent applies a force to a point on an object, at that point
there arises an equal and opposite force back upon the agent.  Were
this not the case, the agent would not experience the process of
exerting a force and we would have no basis for mechanics.  The law
of equal and opposite macroscopic contact forces is thus fundamental
both conceptually and perceptually, but it is legitimate to seek
further underlying connections.  In the case of a stationary object
(fixed to the earth, say), the equal and opposite macroscopic forces
can be said to arise in microscopic interatomic forces in the
neighborhood of the point of contact which act to resist
compression.  This can be traced more deeply still to electromagnetic
interactions involving orbital electrons of adjacent atoms or
molecules, etc.

\bigskip A similar experience of equal and opposite forces arises in
the process of accelerating (pushing on) an object.  It is an
experimental fact that to accelerate an object a force must be
applied by an agent and that the agent will thus experience an equal
and opposite force so long as the acceleration continues.  We argue
that this equal and opposite force also has a deeper physical cause,
which at least in part turns out to also be electromagnetic and is specifically due
to the scattering or interaction with ZPF radiation.  We demonstrate
that from the point of view of a  nearby inertial observer there
exists a net energy and momentum flux (Poynting vector) of ZPF
radiation transiting the accelerating object in a direction
necessarily opposite to the acceleration vector.  The scattering
opacity of the object to the transiting flux creates the
back-reaction force customarily called the inertial reaction.  Inertia is
thus, in part, a special kind of electromagnetic drag effect, namely one that
is acceleration-dependent since only in accelerating frames is the
ZPF perceived as asymmetric.   In stationary or uniform-motion frames
the ZPF is perfectly isotropic.

\bigskip As a first step we must examine in precise detail how we
estimate the change of momentum    $\D\p$ implied in Eq. (9a) before
taking the limit.  Assume for concreteness that the massive object of
rest mass $m$ performs hyperbolic motion under the action of the
external agent with corresponding constant proper acceleration $\a$
along the $x$-axis so that $\a = a\hat{x}$ as in Sec. II.  At proper
time $\D\t$ the object is  instantaneously at rest in the inertial
coordinate frame \Idt \ at the point $(c^2/a, 0, 0)$ of that frame. 
Moreover at the object proper time $\t = 0$ (that corresponds to the
time $t_* = 0$ of \Is), the object was instantaneously detected at
rest at the point $(c^2/a, 0, 0)$ of the laboratory inertial frame
\Is \ by the observer located at that point.  After a short lapse of
laboratory time $\Delta t_*  > 0$ that corresponds to the object
proper time $\D\t$, the object is seen, from the viewpoint of \Is, to
have received from the accelerating agent the amount of impulse or
momentum increment $\D\p_*$.  The expression (9a) but as seen in \Is
\ is thus

$$\F_*={d\p_* \over dt_*} = \lim_{\D t_* \ra 0} {\D \p_* \over \D
t_*} .
\eqno(13)$$

\sn
 At the corresponding object proper time $\D\t$, the object is
instantaneously at rest in the comoving inertial frame \Idt.
Consequently the momentum of the object at proper time $\D\t$ and as
viewed in \Idt \ is of course zero.

\bigskip As proposed in [7],
the force of opposition to the accelerating action imposed by the
external agent does not come from the object itself but from the
all-encompassing vacuum which is restrictively represented herein by only the
electromagnetic ZPF.  Our goal is to show that if this is so, the
force of opposition, $\F_r$, in the subrelativistic case is strictly
proportional to the negative of the acceleration, namely to $-\a$, as
in Eq. (12), and can therefore reasonably be interpreted as a contribution to the
inertia of the object.

\bigskip Taking a vacuum-opposition-to-acceleration for granted, but
not yet assuming that it is proportional to $-\a$, it then follows
that if the force in Eq. (9a) accelerates the object and if our
hypothesis is correct, there must, from Newton's third law, be an
opposite matching reaction force due to the ZPF, $\Fzp$, such that
(see Eq. 12):

$$\Fzp=\F_r=-\F .
\eqno(14)$$

\sn The key is to find whether $\Fzp$ will prove, from relativistic
electrodynamics, to be proportional to  $-\a$. Actually, this should
be true as viewed from any inertial frame whatsoever [14].  To
recapitulate the direction of our argument, we assume in Eq. (13)
that, in the sense of Rindler [12], Newton's second law just provides
a definition for the force entity.  Newtonian mechanics starts at
this point, treating Eq. (13) as a postulate of physics.  The equal
and opposite force of the object on the pushing agent required by
Newton's third law is traditionally assumed to be provided by the
innate inertia of the accelerating object.  Anticipating that the
resistance comes instead from the vacuum, we write this equal and
opposite reaction condition explicitly with the superscript ZP in Eq.
(14).  When we compare Eqs. (9), (12) and (14), it follows that if
the accelerating agent by means of the force $\F$ gives to the object
during a time interval $\D\t$ an impulse or change of momentum
$\D\p$, there must be a corresponding impulse (change of momentum)
$\D\pzp$ provided by the ZPF in exactly the same time interval and as
viewed from the same inertial reference frame but in the opposite
direction to $\D\p$ so that 

$$\D\pzp =-\D\p .
\eqno(15)$$

\sn Hence $\D\pzp$ is the matching reactive counter-impulse given by
the ZPF that opposes the impulse $\D\p$ given by the accelerating
agent.  We refer both $\D \pzp$ and $\D \p$ to the same inertial
frame and in this case to the laboratory frame \Is \ and  write

$$\D \pzp_*=-\D \p_* .
\eqno(16).$$

\sn As this momentum change for the object $\D \p_*$ {\it is
calculated with respect to the inertial frame (that conventionally we
call the laboratory frame) \Is \ and not with respect to any other
frame}, (e.g., the  inertial frame \Idt) it is necessary to calculate
the putative ZPF-induced opposing impulse $\D \pzp_*$ with respect to
the {\it same} inertial frame \Is \ (and not with respect to \Idt \
or any other frame). We write

$$\D \pzp_*=\pzp_*(\D t_*)-\pzp_*(0)=\pzp_*(\D t_*) .
\eqno(17)$$

\sn The momentum $\pzp_*(\D t_*)$ is essentially the integral of
$d\pzp_*$ from \Is-frame time $t_* = 0$ to \Is-frame time $t_* =\D
t_*$.  The last equality follows  from symmetry of the ZPF
distribution as viewed in \Is \ that leads to

$$\pzp_*(0)=0 .
\eqno(18)$$

In what follows we seek  to find a mathematical expression for  the
ZPF-induced inertia reaction force $\Fzp_*$. For this purpose it is
useful to state that from Newton's third law and the force defined
above we can write that the following must be true if our hypothesis
is correct:

$$\lim_{\D t_* \ra 0} {\D \pzp_* \over \D t_*}=
\Fzp_*
= -\F_* =-\lim_{\D t_* \ra 0} {\D
\p_* \over \D t_*} .
\eqno(19)$$

\sn If  the inertia origin propounded here is correct then Eq. (19),
at least in the subrelativistic case, should yield a nonvanishing
force $\Fzp_*$  that is parallel to the direction of the acceleration
$\a = a\hat{x}$, opposite to it, and proportional to the acceleration
magnitude $a = |\a|$.

\vfill\eject
\centerline{\bf IV. INERTIA  REACTION  FORCE  AND THE  ZPF MOMENTUM 
DENSITY}

\bigskip We use Eq. (19) both to evaluate and define the effect that
we will identify with the ZPF inertia force $\Fzp_*$. We concern
ourselves in this section with the ZPF momentum flux entering the
object. Next, in Sec. V, we further develop this analysis (with the
help of Appendix A) to the point of deriving the acceleration-dependent $\Fzp_*$.
Finally in Appendix B we analyze the momentum content.

\bigskip In order to fully grasp the situation we consider the
following simple fluid analogy involving as a heuristic device a
constant velocity and a spatially varying density in place of the
usual hyperbolic motion through a uniform vacuum medium. Let a small
geometric figure of a fixed proper volume $V_0$ move uniformly with constant
subrelativistic velocity ${\bf v}$ along the $x$-direction.  The
volume $V_0$ we imagine as always immersed in a fluid that is
isotropic, homogeneous and at rest, except such that its density
$\rho(x)$ increases in the $x$-direction but is uniform in the $y$-
and $z$-directions.  Hence, as this small fixed volume $V_0$ moves in
the $x$-direction, the mass enclosed in its volume, $V_0 \rho(x)$,
increases.  In an inertial frame at rest with respect to the
geometric figure the mass of the volume, $V_0 \rho(x)$, is seen to
grow.  Concomitantly it is realized that the volume $V_0$ is sweeping
through the fluid and that this  $V_0 \rho(x)$ mass grows because
there is a net influx of mass coming into $V_0$ in a direction
opposite to the direction of the velocity.  In an analogous fashion,
for the more complex situation envisaged in this paper,
simultaneously with the steady growth of the ZPF momentum contained
within the volume of the object discussed above, the object is
sweeping through the ZPF of the \Is \ inertial observer and for him
there is a net influx of  momentum density coming from the background
into the object and in a direction opposite to that of the velocity
of the object.

\bigskip As it is the ZPF radiation background of \Is \ in the act of
being swept through by the object which we are calculating now, we
fix our attention on a fixed point of \Is, say the point of the
observer at $(c^2/a, 0, 0)$ of \Is, that momentarily coincides with
the object at the object proper time $\t= 0$, and consider that point
as referred to the inertial frame \It \ that instantaneously will
coincide with the object at a future generalized  object  proper time
$\t>0$.  Hence  we  compute the \It-frame Poynting vector, but
evaluated at the $(c^2/a, 0,0)$ space point of the \Is \ inertial
frame, namely in \It \ at the  \It \ space-time point:

$$ct_{\t}={c^2 \over a}\snh ,
\eqno(20)$$
$$ x_{\t}=-{c^2 \over a}\csh,
\qquad y_{\t}=0,
\qquad z_{\t}=0.
\eqno(21)$$

\sn This Poynting vector we shall denote by $\Nzp_*$. Everything
however is ultimately referred to the \Is \ inertial frame as that is
the frame of the observer that looks at the object and whose ZPF
background the moving object is sweeping through.  In  order to
accomplish this we first compute

$$\eqalignno{
\< \Ezp_{\t}(0,\t) \times \Bzp_{\t}(0,\t) \> _x &=
 \< E_{y\t}B_{z\t}-E_{z\t}B_{y\t} \> \cr &= \gt^2 \< (E_{y*}-\bt
B_{z*})(B_{z*}-\bt E_{y*}) -(E_{z*}+\bt B_{y*})(B_{y*}+\bt E_{z*}) \>
\cr &=-\gt^2 \bt \< E_{y*}^2+B_{z*}^2+E_{z*}^2+B_{y*}^2 \>  +
\gt^2(1+\bt^2) \< E_{y*}B_{z*}-E_{z*}B_{y*} \> \cr &=-\gt^2 \bt \<
E_{y*}^2+B_{z*}^2+E_{z*}^2+B_{y*}^2 \> & (22)} $$

\sn that we use in the evaluation of the Poynting vector [15]

$$\Nzp_*={c \over 4\pi} <  \Ezp_{\t} \times \Bzp_{\t} >_* = \hat{x}
{c \over 4\pi} <  \Ezp_{\t} (0,\t) \times \Bzp_{\t} (0,\t) >_x .
\eqno(23)$$

\sn The integrals are now taken with respect to the \Is \  ZPF
background (using then the $k$-sphere of  \Is \ introduced in
Appendix C) as that is the background that the \Is -observer
considers the object to be sweeping through.  This is why we denote
this Poynting vector as $\Nzp_*$, with an asterisk subindex instead
of a $\t$ subindex, to indicate that it refers to the ZPF of \Is. 
Observe that in the last equality of Eq. (22) the term proportional
to the $x$-projection of the ordinary ZPF Poynting vector of \Is \
vanishes.  The net amount of momentum of the background  the  object 
has  swept  through  after  a  time  $t_*$,  as  judged  again  from 
the  \Is-frame viewpoint, is

$$\pzp_* = \gzp_* V_* = {\Nzp_*  \over  c^2} V_* = -\hat{x}{1 \over
c^2}{c \over 4 \pi} \gt^2 \bt  {2 \over 3} \<  \E_*^2 + \B_*^2  \>
V_* ,
\eqno(24)$$

\sn which is the complement and clear counterpart of Eq. (B8) of
Appendix B, i.e., the negative of the expression for $\p_*$ evaluated
in Eq. (B9).  Furthermore by means of Eq. (19) we will calculate the
force $\Fzp_*$ directly from the expression for $\pzp_*$ .  These
steps are presented in the next Section.  Prior to  that however we
present a discussion of the conceptual origin of the momentum flux
expression of Eq. (24) complemented with a more detailed derivation
of the cross product of Eq. (23) that is performed in Appendix A.

\vfill\eject
\centerline{\bf V. MOMENTUM FLUX AND NEWTONIAN INERTIA}

\bigskip Any observer at rest in an inertial frame sees the ZPF
isotropically distributed around himself.  The Poynting vector $\Nzp$
and the momentum density  $\gzp=\Nzp/c^2$ of such ZPF vanish for that
observer.  This is of course the case for the observer at rest in
\Is.  Consider now another inertial observer located at a geometric
point that, with respect to  \Is, moves uniformly with {\it constant}
velocity,  $\v= \hat{x} v_x = \hat{x} \b c$.  Imagine the instant of
time when the geometric point is passing and in the immediate
neighborhood of the stationary \Is \ observer.  Both observers
necessarily see the ZPF symmetrically and isotropically distributed
around themselves in their own frames.  However, the ZPF for each
observer is not, because of the Doppler shifts, isotropically
distributed with respect to the other frame.  In the terminology of
Appendix C the \Is-observer is located at the center of his own
$k$-sphere, but the moving point is necessarily located off-center of
the \Is-observer's $k$-sphere. Hence, for  the \Is-observer the ZPF
Poynting vector,  $\Nzp_*$, and the corresponding momentum density,
$\gzp_*$, impinging on the moving point should appear to be
non-vanishing.  Furthermore, because the motion of the geometric
point is uniform, not hyperbolic, both the $\Nzp_*$ and $\gzp_*$ at
the moving geometric point appear to the \Is-observer to be
time-independent constants of the motion.

\bigskip Extend now the consideration above to all the points inside
a small 
$\epsilon$-neighborhood of the previous geometric point that comove
with {\it constant} velocity $\v= \hat{x} c \b$.  Let $V_0$ be the
proper volume of that neighborhood.  Because of length contraction
such neighborhood has, in \Is, the volume $V_* = V_0/\gamma$. 
Clearly to the observer in \Is \ the neighborhood's $\gzp_*$ and 
$\Nzp_*$ do not appear as vanishing because of the uniform motion
with constant velocity,  $\v= \hat{x}\b c$, inducing Doppler shifts
of all the neighborhood's points with respect to \Is.   If the said
neighborhood exactly coincides with the location and geometry of  a 
moving  object of  proper volume  $V_0$  and  rest  mass $m _0$ that
has  the neighborhood's central geometric point at its center, then
according to ordinary mechanics, the object appears to the observer
in \Is \ as carrying a mechanical momentum  $\p_*=\gamma m_0 \v$.

\bigskip We turn now to the object's corresponding ZPF momentum. 
Because the object occupies its proper volume $V_0$ and coincides
with the uniformly moving 
$\epsilon$-neighborhood, it has for the observer at rest in \Is \ an
amount of ZPF momentum, $V_* \g_*=(V_0/\gamma)\g_*$, as described
above.  We re-emphasize that when measured and from the point of view
of the {\it inertial observer comoving with the object}, both the
object momentum and the Poynting vector of the ZPF do exactly vanish,
the last because in $k$-space the object is at the center of
that observer's $k$-sphere (Appendix C).  In the present case of a
constant velocity and zero acceleration for the object, as opposed to
the general case we have been considering of accelerated hyperbolic
motion, the momenta $\p_*$ and $\pzp_*$ above  are both of course
constants. Hence their time derivatives in Eq. (19) both vanish.  We
return to our original hyperbolic motion problem.

\bigskip Let us go back to the paragraph immediately preceding Eq.
(22).  We again compute Eqs. (23) and (24) but perform (23) in more
detail.  From Appendix A, we can compute the Poynting vector of  Eq.
(A4)  that the  radiation should  have  at  the  $(c^2/a,0,0)$ 
point  of  \It \  but referred to \Is \ with the coordinates of Eq.
(22), viz,

$$\eqalignno{
\Nzp_*(\t) &={c \over 4\pi} \< \Ezp \times \Bzp \> \cr &=\hat{x}{c
\over 4\pi} \< E_y B_z - E_z B_y \>  \cr &=-\hat{x}{c \over
4\pi}{8\pi \over 3}\snx \int \A d\w &(25)}$$

\sn where $\Ezp$ and $\Bzp$ stand for $\Ezp_{\t}(0,\t)$  and
$\Bzp_{\t}(0,\t)$ respectively as in the case of Eq. (23)  and where
as in Eqs. (22), (23) and (24) the integration is understood to
proceed over the $k$-sphere of \Is.  The object now is not in uniform
but instead in accelerated motion.  If suddenly at proper time $\t$
the motion were to switch from hyperbolic back to uniform because the
accelerating action disappeared, we would  just need to replace in
Eq. (25) the constant rapidity $s$ at that instant for $a\t$, and
$\bt$ in Eq. (1) would then become $\tanh(s/c)$.  (But then $\Nzp$
would cease to be, for all times onward, a function of $\t$ and
force  expressions as Eq. (28) below would vanish.)  Observe that we
make explicit the $\t$ dependence of this as well as of  the
subsequent quantities below.  $\Nzp_*(\t)$ represents energy flux,
i.e., energy per unit area and per unit time in the $x$-direction. 
It also implies a parallel, $x$-directed momentum density, i.e.,
field momentum per unit volume incoming towards the object position,
$(c^2/a, 0,0)$ of $S$, at object proper time $\t$ and as estimated
from the viewpoint of \Is.  Explicitly such momentum density is 

$$\gzp_*(\t)={\Nzp_*(\t) \over c^2} = -\hat{x} {8\pi \over 3}{1 \over
4\pi c} \snx \int \ew \A d\w ,
\eqno(26)$$

\sn where we now introduce the henceforth frequency-dependent
coupling coefficient, $0 \le \ew \le 1$, that quantifies the fraction
of  absorption or scattering at each frequency.  Let $V_0$ be the
proper volume of the object, namely the volume that the object has in
the reference frame \It \ where it is instantaneously at rest at
proper time $\t$.  From the viewpoint of \Is, however, such volume is
then 
$V_*=V_0/\gt$ because of Lorentz contraction.  The amount of momentum
due to the radiation inside the volume of the object according to
\Is, i.e., the radiation momentum in the volume of the object viewed
at the laboratory is

$$\pzp_*(\t)=V_* \gzp_*={V_0 \over \gt} \gzp_*(\t)
 = -\hat{x}{4V_0 \over 3} c \bt \gt \left[ {1 \over c^2} \int \ew \A
d\w \right] , \eqno(27).$$

\sn which is again Eq. (24).

\bigskip At proper time $\t= 0$, the $(c^2/a,0,0)$ point of the
laboratory inertial system \Is \ instantaneously coincides and
comoves with the object point of the Rindler frame $S$ in which the
object is fixed.  The observer located at $x_* = c^2/a, \ y_* = 0, \
z_* = 0$ instantaneously, at $t_* =0$, coincides and comoves with the
object but because the latter is accelerated with constant
acceleration $\a$, the object {\it according to} \Is \ should receive
a time rate of change of incoming ZPF momentum of the form:

$${d\pzp_* \over dt_*} = {1 \over \gt}{d\pzp_* \over d\t} \bigg|
_{\t=0} .
\eqno(28)$$

We {\it postulate} that such rate of change may be identified with a
force from the ZPF on the object.  Such interpretation, intuitively
at least, looks extremely natural. In this respect Rindler [12] in
introducing Newton's second law makes the following important
epistemological point: ``This is only `half' a law; for it is a mere
definition of force,'' and this is precisely the sense in which we
introduce it here as a definition of the force of reaction by the
ZPF.  If the object has a proper volume $V_0$, the force exerted on
the object by the radiation from the ZPF as seen in \Is \ at $t_* =0$
is then

$${d\pzp_* \over dt_*}=\Fzp_* =-\left[ {4 \over 3}{V_0 \over c^2}
\int \ew \A d\w \right] \a .
\eqno(29)$$

\sn Furthermore

$$m_i= \left[ {V_0 \over c^2} \int \ew \A d\w \right] 
\eqno(30)$$

\sn is an invariant scalar with the dimension of mass.  The expression for $m_i$
differs considerably from the corresponding one in [7] because here, on purpose,
no interaction features were included in the analysis. Such ZPF-particle
interactions will be taken up in future work. Observe that
in Eq. (30) we have neglected a factor of 4/3.  Such factor must be
neglected because a fully covariant analysis (Appendix D) will show
that it disappears.  The corresponding form of $m_i$ as written (and
without the 4/3 factor) is then susceptible of a very natural
interpretation: Inertial mass of an object is that fraction of the 
energy of the ZPF radiation enclosed within the object that
interacts with it (parametrized by the $\ew$ factor in the integrand).  Further
discussion of this point we leave for the Appendices B and D.  Clearly if the
acceleration  suddenly ceases at proper time $\t$, Eqs. (28) and (29)
identically vanish signaling the fact that acceleration is the reason
that the vacuum produces the opposition that we identify with the
force of reaction known as inertia.  From the proper time instant
$\t$ when the acceleration $\a$ is turned off, the object continues
in uniform motion.  The object proceeds onwards with the rapidity $s$
it acquired up to that point, namely $a\t$.  Thus $\bt$ in Eq. (1)
and all quantities from Eqs. (25) to (27) become constants, as the
rapidity $s$ ceases to depend on the proper time $\t$.  Because of
the Lorentz invariance of the ZPF energy density spectrum [16], the
object is left at rest in the inertial frame \It \ and at the center
of the $k$-sphere of the \It \ observer but off-center of the
$k$-sphere of the \Is \ one (Appendix C).  From the \Is \ perspective
the object appears to possess a momentum (which reflects the ZPF
momentum inside $V_0$, a point that will become clear in Appendix
B).  Observe furthermore that in Eq. (30) and previous equations some
cut-off procedure is implicit in that $\ew$ subsides at high
frequencies. (Otherwise we recall the cut-off referred to in Appendix
A according to the prescription of [17].)

\vfill\eject
\centerline{\bf VI. RELATIVISTIC FORCE EXPRESSION}

\bigskip	 The coefficient $m_i$ that we identify with the ZPF contribution
to inertial mass, corresponds then just to the ZPF-induced part of the rest mass of
the object. In order to simplify the discussion that follows, however, we will
usually take this ZPF contribution as all of inertial mass. If the vacuum exerts an
opposition force on the accelerated object of magnitude
$-m_i\a$ as in Eq. (29) and if Newton's third law (Eq. (14)) holds,
then the accelerating agent must exert an active  force $\F$  of
amount  $\F= m_i\a$ to produce the acceleration.  This is Newton's
equation of motion.  The radiative opposition made by the vacuum
precisely coincides time-wise with the onset of acceleration at every
point throughout the interior of the accelerated object, continues
exactly so long as the acceleration persists and is in direct
proportion to the amount of mass associated with that small region. 
Herein lies the intuitive power of the approach.  The only other
alternative is the traditional one of Newton that assumes such 
inertial opposition comes from the object  itself  ``because it has a
fundamental property called mass.'' This however leaves the origin of
inertia unexplained.  Inertia we argue is a phenomenon created by the
vacuum as in Eq. (30). It is important to add that our analysis
yields not just the nonrelativistic Newtonian case but it also
embodies a fully relativistic description within special relativity
[12] at least for the case of longitudinal forces, i.e., forces
parallel to the direction of motion (See however Sec. VII).

\bigskip From the definition of the momentum $\pzp_*$ in Eq. (27),
from Eqs. (28), (29), and the force equation (14) it immediately
follows that the momentum of the object is

$$\p_*=m_i \gt \vec{\beta}_{\t} c ,
\eqno(31)$$

\sn in exact agreement with the momentum expression for a moving
object in special relativity [12].  The expression for the space
vector component of the four-force [12] is then

$$\FF_*=\gt {d\p_* \over dt_*} = {d\p_* \over d\t} ,
\eqno(32)$$

\sn and as the force is pure in the sense of Rindler [12], the
correct form for the four-force immediately follows (recalling (12)
for the space part):

$${\cal F}={d{\cal P} \over d\t} = {d \over d\t}(\gt m_i c, \p) = \gt
\left( {1 \over c}{dE \over dt}, \F \right) = \gt \left( \F \cdot
\vec{ \beta}_{\t}, \F \right) = \left( \FF \cdot \vec{\beta}_{\t},
\FF \right) ,
\eqno(33)$$

\sn in the ordinary way anticipated above.  Consistency with Special
Relativity is established. (For a detailed exposition pertaining to
Eqs. 31--33 see Appendix D.) In Eq. (33)  we have dropped the *
subscript notation for generality.

\vfill\eject
\centerline{\bf VII. GENERAL MOTION}

\bigskip Our analysis so far has been restricted to the case of
simple hyperbolic motion, i.e., rectilinear motion with uniformly
constant proper acceleration $\a$ that remains the same throughout
the trajectory of the object.  Here we give reasons suggesting an
extension of the argument to the case of general motion, i.e., a
motion along a nonrestricted trajectory where the proper
acceleration, henceforth denoted as $\a(\t)$, does not remain
constant, neither in magnitude nor in direction, but on the contrary
changes from one instant to the next of object instantaneous proper
time and becomes thereby a function of  $\t$.  We observe that in our
derivation of Eq. (33) the subrelativistic result $\F=m_i \a$ 
depended only on the instantaneous value of the acceleration $\a$,
i.e., on the hyperbolic moving object instantaneous proper
acceleration, and not in any way whatsoever on the history of the
object motion.

\bigskip For the case of general motion mentioned above, let us
consider a second identical object also performing hyperbolic motion
but with a  trajectory that has exactly the same constant proper
acceleration, $\a_0$,  as the  general motion object at one instant
of the first's proper time $\t=\t_0$.  We thus write
$\a(\t_0)=\a_0$.  But for the second object, i.e., that undergoing
hyperbolic motion, we have shown that Eq. (33) holds and  hence 
that  the accelerating agent must be applying a force $m_i \a_0=m_i
\a(\t_0)$.  Moreover, as argued above, this expression displays
complete independence from the past history of the object's motion,
i.e., expression (33) and related previous expressions are
memoryless.  It is then quite reasonable to assume that as the force
(in the frame that instantaneously comoves with the object performing
hyperbolic motion) is such that it is exactly proportional to the
acceleration and the result is indeed independent of the previous
history of the object motion, the same result should hold for the
identical object in the case of general motion.  This means that for
the case of general motion it should instantaneously hold that

$$\F(\t)=m_i \a(\t) .
\eqno(34)$$

\sn The instantaneous proper force is equal to $m_i$ times the
instantaneous proper acceleration and both are collinear vectors in
the same direction as was  the special case of the hyperbolically
moving object.  What will be required for the confirmation of this
argument however are detailed calculations of several concrete
examples of nonhyperbolic but accelerated motion;  e.g., the case of 
ordinary circular motion with constant angular velocity that yields a
centripetal  acceleration $\a(\t)$ of  constant magnitude.

\vfill\eject
\centerline{\bf VIII. OUTLOOK AND CLOSING REMARKS}

\bigskip The dynamical approach of [7] required mathematical steps
and approximations that led to a certain complexity of presentation. 
The new development here is simpler in that it does not deal with the
dynamics, but exclusively with the form of the ZPF in relation to an
accelerated object.  The final result is derived using standard
relativistic field transformations and does not involve
approximations. A fully covariant analysis is presented in Appendix D.

\bigskip The viewpoint of SED has been taken and matter has been
assumed to be exclusively made of electromagnetically interacting
particles or entities.  Neutrons (presumed to consist of three
charged quarks plus several neutral gluons) are polarizable
electromagnetically-interacting particles displaying at least several
experimentally known  multipole moments.  Neutrons are thus in
principle treatable by our model whereas neutrinos are presumably
strictly neutral and not polarizable.  It is however not clear if
neutrinos do actually have a rest mass and consequently display
subrelativistic behavior or if instead they have no rest mass and
consequently are strictly relativistic.  In either case neutrinos lie
beyond the scope of our present considerations as they do not
interact electromagnetically.  An explanation for the inertia of
non-electromagnetically interacting particles and fields, such as
neutrinos, may follow similar lines in a more sophisticated
development.  If as proposed in this paper (and in [7]) inertia
appears because of the opposition of the vacuum (the electromagnetic
ZPF in the case of SED) to the motion through it of accelerated
particles that interact with its fields ($\Ezp$ and $\Bzp$ fields in
the case of the ZPF of SED), one can naturally conjecture that the
same happens, in a more general way with the zero-point fluctuations
of other fields (like those of the weak and of the strong
interactions) that oppose the accelerated motion through them of
accelerated entities capable of interacting with them.  The general
idea is that rather than postulating an {\it ad hoc} mass-giving
field on top of all the other fields, to examine
instead if inertia can be explained by means of the already
well-established (vacuum) fields, as e.g. the approach of Vigier [1].

\bigskip There is one further conjecture that the present work
suggests.  The four-momentum that the accelerating agent transmits to
the object during the acceleration process should go directly to the
surrounding vacuum field [14].  The question is then whether the
corresponding transferred linear momentum and associated
translational kinetic energy should be radiated away to infinity in
the ordinary manner of electromagnetic {\it radiation} fields or if
instead they stay around in the immediate vicinity of the object in
the usual way of {\it bound} fields [15].  The arguments of Sections
III to VII point rather in the direction of this latter (second)
possibility.  We conjecture that no radiated four-momentum (to
infinity) is produced.  Presumably then only bound or velocity
fields, in the manner of spatially rapidly decaying evanescent fields
outside the accelerating object, are to be expected. As the moving
object, made of electromagnetically interacting matter, enters a
given space region, the electromagnetic modes structure in that
region is modified accordingly. But then the ZPF is thereby also
changed in that region. For this to be possible it may also help to
realize that at the extremely high frequencies involved [7] a strong
nonlinear coupling between matter and ZPF exists.  Non-linearities in
field equations at very high energies are a commonplace theoretical 
possibility [18], e.g., in our case of the simple electromagnetic
field, a generalized form of the equations of classical
electrodynamics may turn out to display non-linearities at very high
energies.  Ordinary Maxwell-Lorentz type equations would just then
represent the linear version of more general nonlinear equations
applicable at all frequencies.

\bigskip  The above view may suggest that molecules, atoms and even
simple electromagnetically interacting particles, create around
themselves and in their immediate vicinity some bound and
nondissipative solitonic-like waves that accompany those particles
wherever they go.  This  speculative view is not new.   Proponents of
source theories have for a long time [19] taken for granted
evanescent fields that surround material particles.  These 
``source-fields'' are presumably able to theoretically substitute for
the action of the more mundane electromagnetic vacuum of  QED in
derivations of ``vacuum'' effects like the Casimir force, the Van der
Waals forces, etc. [19,20].

\bigskip Last but not least we comment on where the concept presented
here fits within the context of ordinary theory and in particular of
standard classical theory.  By looking at what we assumed in the
course of this development we can easily see what the concept
presented here does affect and also {\it what it does not}.  We very
explicitly used the ordinary notion of what force is.  So we cannot
claim any direct explanation of that concept, not even a
clarification of what force means.  With respect to this classical
force concept what we believe we have done is the following. 
Newton's third law requires that the motive force defined in the
second law be counterbalanced by a reaction force.  This has
traditionally been satisfied implicitly by assuming the existence of
inertia of matter.  We propose to have found an explicit origin for
this reaction force, viz. the acceleration-dependent scattering of
ZPF radiation that the accelerated object is forced to move into. 
Our analysis presupposed electrodynamics and special relativity and
other aspects of ordinary classical theory: Electrodynamics and some
aspects of special relativity have been used in our developments
since we used SED (that besides Maxwell's equations also presupposes
the Lorentz force).  As far as radiation reaction is concerned we
merely suspect that it is somewhat connected with the developments
here (and possibly also those in [7] and/or [10]) but so far this is
only a suspicion.  As to what we perceive as the core result of this
work, we claim that our development represents a first step in the
direction of clarifying the origin of what constitutes the essence
of  inertial mass. But it surely falls short in explicating all aspects of the origin
of inertia. Indeed the analysis deliberately excluded any local consideration of
details on particle-radiation interactions and only the electromagnetic aspect of
the physical vacuum was involved.  Only when our approach here can be implemented
by  more detailed models and when it can be extended to other
components of the vacuum (other vacuum fields, e.g., weak, strong
interactions) will inertial mass be able to be made assimilable to a
property material objects of all sorts have because material objects
affect the structure of the underlying modes of the vacuum fields in
which they are permanently immersed.

\bigskip Finally we make two disclaimers. First, our development was
exclusively made within the context of classical theory (SED), so it
is too early to say much  about connections with quantum theory
[21]. There for example remains the more technical question of how the SED
renormalization procedure proposed herein (Appendix C) matches with its
QED
counterpart as both lead to essentially the same result.  Second, we are not
prepared to face the issue of how and in what sense our development might possibly
affect or relate to general relativity (beyond what was briefly mentioned in
Section I concerning Sakharov's hypothesis). 

\bigskip
\centerline{\bf ACKNOWLEDGMENTS}

\bigskip
A. R. thankfully acknowledges
detailed and extensive correspondence
with Dr. D. C. Cole that was instrumental in clarifying or
developing various arguments in this article.
B.H. wishes to thank Prof. J. Tr\"umper and
the Max-Planck-Insititut f\"ur Extraterrestrische Physik for
hospitality during several stays. 
We acknowledge support of this work by
NASA contract NASW-5050.

\vfill\eject
\centerline{\bf APPENDIX A:}
\centerline{\bf ZPF-AVERAGED PRODUCTS FOR THE ZPF POYNTING VECTOR}

\bigskip Here we proceed with the explicit evaluation of the averaged
products of the electric and magnetic field components $<E_i  B_j >;
\ i, j = x, y, z$, that enter in the expression for the ZPF Poynting
vector corresponding to the radiation being swept through by the
accelerated object as calculated from the viewpoint of the observer
at rest at $(c^2/a, 0, 0) $ of \Is.    The resulting averaged
products are part of the more extensive presentation of the approach
implemented in Section V. (Note that for simplicity we generally drop
the ZP-superscript notation for $\E$ and $\B$.)

\bigskip We will see that the diagonal ones $<E_i  B_i >; \ i= x, y,
z$, all vanish necessarily leading to the consequence that

$$<\E \cdot \B>=0
\eqno(A1)$$

\sn as was to be expected. Of the remaining $<E_i  B_j >$ averaged
products we shall show that those products involving one component in
the direction of the acceleration $\a$, i.e., one $x$-component of a
field, $E_x$ or $B_x$, vanish irrespectively of the other component: 

$$<E_x  B_j >=0;  \ j = x, y, z
\eqno(A2)$$

\sn and 

$$<E_i  B_x >=0; \ i = x, y, z .
\eqno(A3)$$

\sn The only products that are spared are $< E_y B_z >$ and $< E_z
B_y >$.  They, moreover, will turn out to be the opposite of each
other (see below) so that the momentum density (Eq. [15] ) becomes

$$\gzp_*={\Nzp_* \over c^2}={1 \over c^2} {c \over 4\pi} \< \Ezp
\times \Bzp \> =\hat{x}{1 \over c^2} {c \over 4\pi} \< E_y B_z - E_z
B_y \> =\hat{x}{1 \over c^2} {c \over 2\pi} < E_y B_z> .
\eqno(A4)$$

\sn This corresponds then to the \Is \ fluid swept by the motion of
the object across \Is \  according to the observer at rest in \Is. 
We only compute the Poynting vector $\Nzp_*$ that should be
understood in the same sense and exactly as defined in Sec. IV.
Explicit calculations evaluating Eqs. (A1--A4) follow below.  In all
those calculations, we do the averaging over the phases with

$$\displaylines{
\bigg\langle
\cos \[ k_x{c^2 \over a} \csh - {\w c \over a} \ \snh - \theta({\bf
k},\l) \] \hfill \cr
\hfill \cdot
\cos \[ k'_x{c^2 \over a} \csh - {\w' c \over a} \ \snh - \theta({\bf
k}',\l') \]
\bigg\rangle} $$
$$ =\hf \delta_{\l\l'}\delta({\bf k}-{\bf k}') .
\eqno(A5)$$

\sn Notice that this equation is not properly normalized.  A
normalization factor $(2\pi)^3/V$, where $V$ is the so called
electromagnetic cavity volume, has been omitted for brevity on the
right hand side.  A corresponding compensating normalization factor
$(V/(2\pi)^3)^{1/2}$  has also been omitted from the expressions for
the fields, starting from Eqs. (3) and (4) above.  These
normalization factors, standard
 in physical optics, have been introduced also to SED [13], but as
they mutually cancel after phase averaging we do not make them
explicit in this paper.

\bigskip After using Eqs. (10) and (11) and phase averaging with
(A5), we find the first component for the case $i = x, \ j = x$:

$$< E_x B_x > =\hf  \Sm \Ik \Hw \he_x \ke_x 
\eqno(A6)$$

\sn and as

$$\Sm \he_x \ke_x = 0
\eqno(A7)$$

\sn in partial confirmation of Eq. (A3),

$$<E_x B_x >=0 .
\eqno(A8)$$

\sn Note that we perform all $k$-integrations in the $k$-coordinates
of \Is, centered in the $k$-origin of \Is \  as stated in Sections IV
and V.  In an analogous way to Eq. (A6), we easily find after
averaging with Eq. (A5) that

$$<E_x B_y>= \hf \Sm \Ik \Hw \he_x \csh \[ \ke_y + \tnh \he_z \]  ,
\eqno(A9)$$

\sn and as

$$\Sm \he_x \ke_y = {k_z \over k} = \hk_z ,
\eqno(A10)$$

$$\Sm \he_x \he_z = - \hk_x \hk_z ,
\eqno(A11)$$

\sn and because of the angle integrations in
$\Ik \dots = 
\int k^2 dk \int d\Omega \dots =
\int k^2 dk \int \sin \theta d \theta \int d \phi \dots$,

$$\int \hk_z d\Omega =0 ,
\eqno(A12)$$

$$\int \hk_x \hk_z d\Omega = 0 , 
\eqno(A13)$$

\sn we can confirm that as stated in Eqs. (A2) and (A3)

$$< E_x B_y > = 0 .
\eqno(A14)$$

\bigskip As the problem is symmetric around the acceleration
direction, i.e., the $x$-axis, it follows that if $< E_x B_y >=0$,
then also

$$< E_x B_z > = 0 .
\eqno(A15)$$

\sn         For the $yx$-component, after phase averaging, we find    

$$<E_y B_x>= \hf \Sm \Ik \Hw \ke_x \csh \[ \he_y - \tnh \ke_z \]  .
\eqno(A16)$$

\sn         Furthermore

$$\Sm \he_y \ke_x = - \hk_z 
\eqno(A17)$$

\sn         and

$$\Sm \ke_x \ke_z = -\hk_x \hk_z 
\eqno(A18)$$

\sn         that yield, as in Eq. (A8), vanishing solid angle
integrations.  We obtain

$$< E_y B_x > = 0 .
\eqno(A19)$$

\sn         In analogous fashion:

$$<E_y B_y>= \hf \Sm \Ik \Hw \cxh \[ \he_y - \tnh \ke_z \] \[ \ke_y +
\tnh \he_z \] ,
\eqno(A20)$$

\sn         for which we use the identities

$$\Sm \he_y \ke_y = 0 ,
\eqno(A21)$$

$$\Sm \he_z \ke_z = 0 ,
\eqno(A22)$$

$$\Sm \he_y \he_z = -\hk_y \hk_z ,
\eqno(A23)$$

\sn         and

$$\Sm \ke_y \ke_z = - \hk_y \hk_z
\eqno(A24)$$

\sn         to show that

$$< E_y B_y > = \hf \Sm \Ik \Hw \cxh \tnh (-\hk_y \hk_z + \hk_y
\hk_z) = 0 . \eqno(A25)$$

\sn         In a similar way and after phase averaging, it follows
that

$$< E_y B_z > = \hf \Sm \Ik \Hw \cxh \[ \he_y-\tnh \ke_z \] \[ \ke_z
- \tnh \he_y \] \eqno(A26)$$

\sn         where we use

$$\Sm \he_y \ke_z = \hk_x ,
\eqno(A27)$$

$$\Sm \ke_z \ke_z = 1 - \hk_z^2 
\eqno(A28)$$

\sn         and

$$\Sm \he_y^2 = 1-\hk_y^2 ,
\eqno(A29)$$

\sn         so that

$$< E_y B_z > = \hf \Sm \Ik \Hw \cxh \cdot$$
$$\bigg\{ \[ 1 + \txh \] \hk_x - \tnh (1 - \hk_y^2) - \tnh (1-
\hk_z^2) \bigg\} .
\eqno(A30)$$

\sn         The first term inside the curly brackets, proportional to
$\hk_x$, vanishes after angle integrations.  For the others we use
the fact that $2 - \hk_y^2 - \hk_z^2 = 1 + \hk_x^2$ and after minimal
algebra then

$$<E_y B_z > = -{1 \over 4} \snx \int d\w \A \int d\Omega (1+\hk_x^2)
\eqno(A31)$$

\sn         where the divergence of the integration may be damped by
a well-known convergence form factor (or more roughly a frequency
cut-off) that we do not need to make explicit at this stage.  Such a
feature physically represents a frequency beyond which no material
subparticle however small is going to be able to react to the
radiation [17].  This is not a cut-off in the ZPF itself but is
introduced because wavelengths smaller than the size of the minimal
relevant entities, say partons [7], cannot produce any translational
interactions but only presumably internal deformation of the
``parton'' [17].  We thus obtain

$$< E_y B_z > = -{4 \pi \over 3} \snx \int \A d\w 
\eqno(A32)$$

\sn         where, as explained above, an implicit cut-off or
convergence form factor [17] in the frequency integration may
henceforth be implemented [7].  For $< E_z B_x >$ we obtain from
symmetry around the $x$-axis and Eq. (A19) that

$$<E_z B_x > = 0 .
\eqno(A33)$$

\sn         The case of $<E_z B_y >$ is done exactly as that for
$<E_y B_z >$ and gives the opposite value

$$\eqalignno{ < E_z B_y > &= \hf \Sm \Ik \Hw \cxh \tnh (2 - \hk_y^2 -
\hk_z^2 ) \cr &= {4 \pi \over 3} \snx \int \A d\w . &(A34)}$$

\sn         Due to the cylindrical symmetry around the $x$-axis, the
case of $<E_z B_z >$, as the one for  $<E_y B_y >$,  must vanish:

$$< E_z B_z > = 0 .
\eqno(A35)$$

\vfill\eject
\centerline{\bf APPENDIX B:}
\centerline{\bf THE ZPF MOMENTUM CONTENT OF AN ACCELERATED OBJECT}

\bigskip The development of this ZPF momentum content approach serves
to complement the ZPF momentum flux  approach of Sections IV and V
but  is {\it totally independent} from it.   Newton's third law as
expressed in Eq. (19) implies that $\Fzp_*$ is  the opposite of 
$\F_*$.  In Section V we calculate  $\Fzp_*$.  Here instead we
calculate $\F_*$. To calculate $\F_*$ we use the fact, also given in
Eq. (19), that $\F_*$ is just the time derivative of $\p_*$.  This
$\p_*$ is not the ZPF momentum flowing through the object that we
identified with $\pzp_*$ above, but rather the momentum contained
within the body that should be due to the ZPF, as will become clear
as the argument progresses. This means that we need to reach an
expression for $\p_*$, the object momentum, in terms of the ZPF and
the particle electromagnetic coupling parameters. 

\bigskip First we develop a connection between $\p_*$ and $\pzp_*$
before defining $\p_*$ more precisely, so we examine the
$\D\pzp_*=-\D\p_*$ condition of Eq. (16) referring everything,
including the ZPF background, to the \Is \ inertial frame.  We need
to estimate the net ZPF momentum, $\D\pzp_*$, and momentum density
that entered into the object in a time interval $\D t_*$ because of
the object's sweeping accelerated motion through the electromagnetic
background (Secs. IV and V).  As we take the integrated impulses 
$\pzp_*$ and $\p_*$   both to be zero at time $t_* = 0$ and they
start to change at the same time, owing to Eq. (19) we should end up
with

$$\pzp_*=-\p_* 
\eqno(B1)$$

\sn where we have integrated $\Fzp_*$ and $\F_*$ over the \Is-frame
time $t_*$.  Moreover, because both $\pzp_*$ and $\p_*$ are referred
to the laboratory inertial frame \Is, then, $\pzp_*=V_*\gzp_*$ and
$\p_*=V_*\g_*$ that with Eq. (B1) lead to 

$$\gzp_* = -\g_* .
\eqno(B2)$$

\sn We have calculated $\pzp_*$ in Secs. IV and V. After confronting
it with the result for  $\p_*$ (which we calculate below), we shall
verify Eq. (B1), as is necessary for a self-consistent interpretation
of momentum.  Therefore, as it will be explicitly confirmed below,
the previous result  for  $\gzp_*$ will be the  negative of  the 
expression (B7)  for  $\g_*$ below, that is calculated here in a
direct and independent way.  Eq.  (B2) also makes sense intuitively
with the aid of the fluid analogy of Sec. IV.  In the short time
interval  $\D t_*$  the  momentum  density  inside  the  object 
increases  in  an  amount   $\D\g_*$.  The corresponding change of 
$\gzp_*$  in the same time interval is $\D\gzp_*$.  Hence if 
$\D\g_*$ is the internal change in momentum per unit volume of the
ZPF in the interior of the object and referred to \Is, then 
$-\D\gzp_*$ is the corresponding  net  amount of momentum per unit
volume that in the same time interval $\D t_*$ is swept inside the
object volume due to the hyperbolic motion of the object through the
ZPF of \Is.  We now calculate the momentum density $\g_*$ in a direct
manner.

\bigskip If the central idea that the ZPF is the entity responsible
for a contribution to the inertia reaction force is valid, then, {\it from the point
of view of the \Is \ frame}, it follows that when the accelerating agent
applies the external force $\F_*$ and accelerates the object,
necessarily the accelerating agent is doing work against the vacuum
fields and hence the  energy provided by the accelerating agent is
going to be stored somewhere in that vacuum.  We envision a
dichotomous situation, i.e.,  two possible alternative ways in which
the vacuum may store this energy and its corresponding momentum
provided by the accelerating agent .  One way is that the energy is
``radiated away'', i.e., the vacuum traveling modes radiate it out
far away from the accelerated body.  The other  is that such energy
and momentum are not indeed radiated out, but on the contrary stay
bound in the manner of  velocity fields [15] in and around the  body
presumably in association to the ZPF electromagnetic bound modes
created by microscopic charge and current structures within the
body.  We propose that the first hypothesis, i.e., that the
four-momentum is radiated away to infinity is less natural than the
second, i.e., than the one that states that the four-momentum remains
bound to the body:  If a billiard ball is hit by the billiard club
(its accelerating agent) receiving this way energy and momentum,
i.e., four-momentum, and if this four-momentum were to escape to
infinity, there would not be a readily natural way to explain how the
ball in its turn subsequently transmitted all of its acquired
four-momentum at the instant of a subsequent collision with another
identical billiard ball that was standing still on the billiard
table.  The original ball's four-momentum had been radiated away to
infinity!  Explanations can probably be concocted but not one that is
in an obvious  way natural.  We assume then only the second
hypothesis in this dichotomy which assumes that the acquired
four-momentum is not stored far away from the accelerated object but,
on the contrary, that it is stored within and in the body's immediate
vicinity. Imagine an idealized body made of a simple electromagnetic
cavity. The modes of the cavity contain ZPF energy and if the cavity
moves they can transport electromagnetic momentum.

\bigskip We therefore, in spite of its still very preliminary nature,
have to follow  this second hypothesis, namely that when a massive
body moves with respect to some inertial frame the moving body drags
with itself  as velocity fields within its ZPF-bound electromagnetic
modes the corresponding translational kinetic energy and momentum the
body has with respect to that frame.  Hence we assume that the
momentum $\p_*$ of the moving object referred to the inertial frame
\Is \ is due to the ZPF that interpenetrates the object.  In order to
find $\p_*$ we first calculate the ZPF momentum density $\g_*$
associated with the exact position of our accelerated object attached
to the frame $S$, at the point $(c^2/a, 0, 0)$ of that frame.  As the
motion is hyperbolic and in the $x$-direction we only need to
calculate

$$\g_* = \hat{x} g_{*x} = {\N_* \over c^2} = \hat{x} {1 \over c^2} {c
\over 4\pi}
\< \Ezp_* (0,\t) \times \Bzp_* (0,\t) \> _x
\eqno(B3)$$

\sn where by this we mean the momentum density $\g_ *$ and the
associated Poynting vector $\N_*=\hat{x}N_{*x}$ [15] due to the ZPF
as measured in the \Is \ frame at the object's space-time point at
object proper time $\t$, namely : 

$$t_*={c \over a} \snh , \ x_*={c^2 \over a} \csh ,	\ y_*=0, \ z_* =
0 , 
\eqno(B4)$$	

\sn of that \Is \ frame.

\bigskip However, as we are calculating the ZPF momentum associated
with the object and the object is instantaneously at rest in the \It
\ frame, the calculation should be done with this in mind.  This
means (Appendix C) that the integrals should be performed over the
$k$-sphere of the \It \ frame.  Additional support for this view
comes from the clear interpretation it yields.  In order to evaluate
Eq. (B3) we need then to compute the relevant averaged cross-product
of the electric and magnetic  fields in \Is \ but evaluated at the
object space-time point of Eq. (B4):
										
$$\eqalignno{
\< \Ezp_* (0,\t) \times \Bzp_* (0,\t) \> _x &=
 \< E_{y*}B_{z*}-E_{z*}B_{y*} \> \cr &= \gt^2 \< (E_{y\t}+\bt
B_{z\t})(B_{z\t}+\bt E_{y\t}) -(E_{z\t}-\bt B_{y\t})(B_{y\t}-\bt
E_{z\t}) \> \cr &=\gt^2 \bt \<
E_{y\t}^2+B_{z\t}^2+E_{z\t}^2+B_{y\t}^2 \>  + \gt^2(1+\bt^2) \<
E_{y\t}B_{z\t}-E_{z\t}B_{y\t} \> \cr &=\gt^2 \bt \<
E_{y\t}^2+B_{z\t}^2+E_{z\t}^2+B_{y\t}^2 \>  & (B5)} $$

\sn where we have made a Lorentz transformation of the fields [15]
from those of \Is \ to those of \It.  One of the resulting terms is
found to be zero because it is proportional to the $x$-component of
$\Nt$, the statistically averaged ZPF Poynting vector as viewed in
the inertial frame \It \ at a point fixed in that frame.  Furthermore

$$\< E_{y\t}^2+B_{z\t}^2+E_{z\t}^2+B_{y\t}^2 \> ={2 \over 3} \<
\E_\t^2 + \B_\t^2 \> ={2 \over 3}	8 \pi U_{\t}=2 \cdot {8 \pi \over
3} \int \Ap d\w' 
\eqno(B6)$$

\sn where $U_{\t}$ means the ZPF energy density in \It \ given by the
frequency integral that, as mentioned above, is carried out over the
$k$-sphere of \It.  In the last step we have simplified the notation
such that hereafter  primes denote quantities referred to \It , and
unprimed quantities mean quantities referred to \Is, so we do not
need to carry everywhere the $\t$ and/or * subindices.  From Eqs.
(B3), (B5) and (B6) then

$$\g_*=\hat{x}{1 \over c^2}{c \over 2 \pi}{8 \pi \over 3} \int \Ap
d\w' \snh \csh . \eqno(B7)$$

\sn From this we can get the ZPF momentum corresponding to the volume
of the object all as referred to the inertial observer of \Is,

$$\p_* = \g_* V_* = \hat{x} {1 \over c^2}{c \over 4 \pi} \gt^2 \bt {2
\over 3}
\< \E_{\t}^2 + \B_{\t}^2 \> V_* ,
\eqno(B8)$$

\sn where $V_*$  is the  volume  that  the  object presents  to the
observer in \Is, namely $V_*=V_0/\gamma$ because of
Lorentz-contraction ($V_0$ is the proper volume of the object).  As
$\hat{x} \bt c$ is the object velocity and $< \E_{\t}^2 + \B_{\t}^2 >
V_0$ is proportional to the proper ZPF energy contained in the volume
of the object, it is a simple matter to realize that Eq. (B8) does
indeed have the form of a relativistic momentum ($\p=\gamma m_0 {\bf
v}$ with $m_0$ the rest mass). 

\bigskip Having established that Eq. (B8) correctly represents the
total ZPF momentum instantaneously contained in the proper volume
$V_0$ of the object in question we next reintroduce a coupling
parameter, $\eta$, which is now a function of frequency, $\ew$, and
which will parametrize the amount of interaction (absorption or
scattering) at every frequency $\w$ between the object and the
radiative momentum flux associated with it.  This will provide us the
effective physical momentum of the object. With this in mind we write:

$$\p_*=\hat{x} {4 V_0 \over 3} c \bt \gt \left[ {1 \over c^2} \int
\ewp \Ap d\w' \right] . \eqno(B9)$$

\sn Recognizing that this is the momentum of the body and that the
accelerating force $\F_*$ applied by the external agent is just the
rate of change of $\p_*$ with time we have $\F_*=d\p_* / dt_*$. From
Eq. (19) (Newton's Third Law) we then obtain the expression for the
force $\Fzp_*=-\F_*$ that the ZPF applies back to the object in
opposition to the external agent's accelerating action

$$\Fzp_*=-\F_* =-{d\p_* \over dt_*} = -{1 \over \gt}{d\p_* \over d\t}
\bigg| _{\t=0}= -\left[ {4 \over 3}{V_0 \over c^2} \int \ewp \Ap d\w'
\right] \a .
\eqno(B10)$$

\sn As already mentioned above, the 4/3 factor becomes unity when a
fully covariant evaluation is performed (Appendix D).  Observe that
Eq. (B10) exactly reproduces the result for the inertia reaction
force of Eq. (29) and its associated inertial mass of Eq. (30). 
Observe furthermore that $m_i$, when the  4/3 factor is obliterated
(Appendix D),  has a clear interpretation, namely it is exactly the
amount of energy of the ZPF that lies inside the object's proper
volume $V_0$ and that does actually interact with it  (as depicted by
the frequency-dependent coupling efficiency function $\ew$ divided by
$c^2$ to give it in units of mass).  According to the view presented
here it is the parameter $V_0$ and the spectral function $\ew$ that
determine the inertial mass of the object.  The expression for $m_i$
would precisely fit the energy of the ZPF (divided by  $c^2$) that
couples to the object and that lies within its volume as properly
assumed in the  approach of this Appendix.  This interpretation is
not so clearly found for the physical assumptions of the  approach of
Sections IV and V and Appendix A, but still in that approach  the
same $m_i$ was found as required.  Interestingly enough this 4/3
factor is the same factor found in the electrodynamics of classical
charged particles [12, 15].  We found here further motivation for
pursuing  the fully covariant analysis of Appendix D.

\vfill\eject
\centerline{\bf APPENDIX C:  DISCUSSION ON THE EVALUATION}
\centerline{\bf OF THE TIME RATE OF CHANGE OF ZPF MOMENTUM}

\bigskip This article has simplicity as one of its objectives. 
Insights however, are gained if we pay the price of going through a
few intricacies.  In our analysis there appears a momentum density
from the ZPF radiation in accelerated frames that is dependent on the
proper time of the accelerated object.  Such proper time may be
removed by means of an artifice.  We believe that doing so would be
physically incorrect and propose that such time dependence actually
has an important physical meaning.  This entails the adoption of a
convention for the evaluation  of otherwise indefinite integrals. 
Removal of the proper time would imply a completely different
situation than the one we are dealing with here.  The momentum
density mentioned above depends on the object proper time.

\bigskip {\bf A. Improper integrals and ``time removal'' technique}

\bigskip First of all we notice that the implicitly unbounded $k$ and
$\w$ integrations of Sections IV, V and Appendices A and B when there
is no $\ew$ (i.e. set $\ew \equiv 1$) and the limits for $k$ or $\w$ are indeed 0
and
$\infty$, constitute what are customarily called improper integrals.  Improper
integrals require some definition as they do not yield a unique
well-defined value.  In Sections IV and V and Appendices A and B we
implicitly adopted a definition.  As the inertia resistance force
expression of Eq. (29) resulted from integrating over the $\w$ or $k$
of \Is \ in an explicit manner and as that result was physically
sound, as follows from the clear delineation of Section III, we
adopted the way of integrating implicit in that procedure as the
actual definition of the improper integral.  That this result is not
necessarily unique is however seen from what follows.  By performing
a change of variable of integration, from that of \Is=\It$(\t=0)$ to
that of \It$(\t>0)$, it can be shown that a different result is
obtained.  Moreover the procedure removes the proper time or
$\t$-dependance of the expressions.  Thus it is sometimes referred to
as a ``time removal procedure'' [22].

\bigskip The procedure leads to a seemingly paradoxical situation as
the integration apparently yields two different values.  The paradox
is easily resolved.  Its solution yields insight into the origin of
$d\pzp_*/dt_*$, the time rate of change of the radiation momentum in
Eq. (28), and on its proposed connection to Newtonian inertia.

\bigskip For the particular SED form of the ``time removal''
procedure [23,24], we start from the expression derived in Appendix A,

$$\displaylines{ <E_y B_z>\equiv<E_y(0,\t) B_z(0,\t)>=\bigg\langle
\Sm \Smp \Ik \Ikp  \hfill \cr
\times \csA \[ \he_y-\tnA \ke_z \] \csA \[ \kep_z-\tnA \hep_y \]
H_{zp}(\w) H_{zp}(\w') \cdot
\hfill (C1) \cr
\cos \[ k_x {c^2 \over a} \csA - {\w  c \over a} \snA - \theta({\bf
k},\l) \]
\cos \[ k_x'{c^2 \over a} \csA - {\w' c \over a} \snA - \theta({\bf
k}',\l') \]
\bigg\rangle , }$$

\sn where we define

$$A \equiv {a\t \over c}
\eqno(C2)$$

\sn and the averaging we perform by means of Eq. (A5).  This yields
again as in Appendix A:

$$< E_y B_z > = \hf \Sm \Ik \Hw \cxA \[ \he_y-\tnA \ke_z \] \[ \ke_z
- \tnA \he_y \] . \eqno(C3)$$

\sn Of course that Eq. (C3) is identical to Eq. (A26) and if we
follow in the ordinary way the command dictated by the summation over
lambda operator and the integration over ${\bf k}$-operator of Eq.
(C3) the result cannot be different from the one of  Eq. (A31).  At
this stage however, one may decide to change the dummy variable of
integration and instead of integrating over the four-vector
components $(k,{\bf k})$ of \Is,  integrate over the four-vector
components $(k',{\bf k}')$ of \It \ by means of the corresponding
transformation  where of course $k'=\w'c$:

$$k=c\w=k'\csA+k_x'\snA=k'\gt+k_x'\bt\gt ,
\eqno(C4a)$$

$$k_x=k_x'\gt+k'\bt\gt=k_x'\csA+k'\snA ,
\eqno(C4b)$$

$$k_y=k_y' \ , \ \ \ k_z=k_z' .
\eqno(C4c)$$

\sn Moreover we know that

$${d^3k \over k}={d^3k' \over k'} .
\eqno(C5)$$

\sn Both integrations, the one over $d^3k$ in the coordinates of
\Is,  and the one over $d^3k'$ in the coordinates of \It, are at
least formally performed over all $k$-space.  Moreover as all space
for $k$ corresponds to all space for $k'$, and furthermore the
Jacobian of the transformation that follows easily from Eqs. (C4a)
and (C5)

$${d^3k \over d^3k'}=J({\bf k},{\bf k}')=\gt \( 1+{k_x' \over \w'}
c\bt \)
\eqno(C6)$$

\sn is, except for $\w'=0$, nonsingular, then the integrations are
both extended over all space whether when done over $k$ or when
performed over $k'$.  Changing the integration variable by Eq. (C5)
and using again Eqs. (A27--A29) we obtain

$$\eqalignno{ <E_y B_z>_R &= \hf \Ikk {\hbar c \over 2 \pi^2} k^2
\cxA \cdot \cr & \qquad \qquad  \cdot \lb \[ 1+\txA \] {k_x \over k}
- \tnA \( 1-{k_y^2 \over k^2} \) - \tnA \( 1-{k_z^2 \over k^2} \) \rb
\cr &= {\hbar c \over (2\pi)^2} \Ikk k' k_x' . &(C7)}$$

\sn The subindex $R$ stresses the difference between the case here
and the one in the body of the paper.  Observe that the proper time
$\t$ and hence $A$ of Eq. (C2) have disappeared from the integral and
this even before performing the integration: the proper time $\t$ has
been ``removed''.  The last equality demanded minor omitted algebra. 
Then as

$$\int k_x' d\Omega'=0
\eqno(C8)$$

\sn the integral in Eq. (C7) vanishes identically and thus

$$<E_y B_z>_R = 0 .
\eqno(C9)$$

\sn Symmetry then yields that also

$$<E_z B_y>_R=0 .
\eqno(C10)$$

\sn It has been a simple matter to establish by this procedure that
the non-zero products of Section IV and Appendix A vanish.  It is
equally simple to establish that all the other products that already
vanished there also vanish here and then that

$$<E_i B_j>_R=0 \ , \ i,j=x,y,z .
\eqno(C11)$$

\sn Consequently according to this viewpoint the Poynting vector and
hence the momentum density $\g_R$ identically vanish

$$c^2 \g_R = \N_R = 0 .
\eqno(C12)$$

\sn This is not so surprising if we propose the view that what we
have done by means of the transformation from $(k,{\bf k})$ to
$(k',{\bf k}')$, i.e., from the variables of \Is \ to the variables
of \It, is to reevaluate the averaged Poynting vector incident on the
moving object not now as the observer at rest in \Is \ estimates it,
in which case such $\N$ does not vanish, but as viewed from the space
point $(c^2/a, 0,0)$ of the inertial frame \It \  where the object is
found at rest at proper time $\t$ and from where the ZPF of \It \ is
necessarily seen as homogeneously and isotropically distributed.

\bigskip Next we show how the formalism of Eq. (C4), that involves a
transformation of the four-wavevector $(k,{\bf k})$ from \Is \ to \It
\ and its corresponding inverse transformation that transforms from
the $(k',{\bf k}')$ four-wavevector of \It \ back into that of \Is,
take us, at least formally, from the Poynting vector $\N$ impinging
on the object as estimated in \Is \ to the same $\N$ but as estimated
in \It, and the second transformation back from the $\N$ impinging on
the object as  viewed in \It \ to the same but as viewed in \Is.

\bigskip From Eq. (A4) the  incident $\N$ on the object, as viewed in
\Is, is $(c/2\pi)$ times the expression $<E_yB_z>$ of Eqs. (A30) and
(C3).  Hence because of Eqs. (C7) and (A4) the $(k,{\bf k}) \ra
(k',{\bf k}')$ transformation of (C4) takes us from the $\N$ of Eq.
(25) to 

$$\N = {\hbar c^2 \over (2\pi)^3} \Ikk k'k_x'
\eqno(C13)$$

\sn that as shown below is the $\N$  at proper time $\t$ impinging on
the object according to an inertial observer instantaneously
coinciding and comoving with the object, hence at rest in \It.  Eq.
(C13) of course vanishes (after the angle integration Eq. (C8)) as it
should because the ZPF in every inertial frame is homogeneous and
isotropic.  Recall that our ultimate proper time of evaluation is
indeed $\t=0$ when \It \ becomes \Is=\It$(\t=0)$.  The corresponding 
$\N(\t)$ at proper time $\t$ exactly equal to zero, $\N(\t=0)$, also
vanishes as follows from Eq. (B6).

\bigskip Next we perform the inverse four-wavevector transformation
$(k',{\bf k}') \ra (k,{\bf k})$ from the four-wavevector of \It \ to
that of \Is \ on the expression that corresponds to the $\N$ of Eq.
(C13), that is the $\N$ viewed in \It.  Nevertheless, we first derive
the explicit form of the $\N$ of \It \ and show that it indeed
corresponds to Eq. (C13) and only then, at a second stage, perform
the aforementioned transformation.

\bigskip Consider the ZPF $\E'$ and $\B'$ fields of \It \ at time
$t_{\t}$ of \It \ and for concreteness at the point $x_{\t}= c^2/a, \
y_{\t} = 0, \ z_{\t} = 0$, that is the point of \It \ that at object
proper time $\t$ coincides and comoves with the object.  Take the
averaged Poynting product of these $\E'$ and $\B'$ fields,

$$\N \bigg| _{I_{\t}} = {c \over 4\pi} <\E' \times \B'> = \hat{x}{c
\over 4\pi} <(\E' \times \B')_x> = \hat{x} <E_y'B_z' - E_z'B_y'> ,
\eqno(C14)$$

\sn where the second equality follows from symmetry. We use the
primed notation for all fields and variables of \It.  From Eqs. (3),
(4) and (5) then

$$\N \bigg| _{I_{\t}} =\hat{x}{c \over 4\pi} \sum_{\l_1' \l_2'} \int
d^3k_1' \int d^3k_2' \ H(\w_1') H(\w_2')
\[ \he_{1y}' \kep_{2z} - \kep_{1y} \he_{2z}' \] \cdot
\eqno(C15)$$
$$
\< \cos \( k_{1x}'x'+k_{1y}'y'+k_{1z}'z'-\w_1't_{\t}+\theta({\bf
k}_1',\l_1') \)
     \cos \( k_{2x}'x'+k_{2y}'y'+k_{2z}'z'-\w_2't_{\t}+\theta({\bf
k}_2',\l_2') \) \> .
$$

\sn Applying the phase averaging of (A5) we have

$$
\< \cos \( k_{1x}'x'+k_{1y}'y'+k_{1z}'z'-\w_1't_{\t}+\theta({\bf
k}_1',\l_1') \)
     \cos \( k_{2x}'x'+k_{2y}'y'+k_{2z}'z'-\w_2't_{\t}+\theta({\bf
k}_2',\l_2') \) \>
$$
$$=\hf \delta_{\l_1' \l_2'} \delta({\bf k}_1'-{\bf k}_2') ,
\eqno(C16)$$

\sn and we get

$$\N \bigg| _{I_{\t}} = \hat{x} {c \over 4\pi} \Smp \Ikp \Hwp \hf [
\hep_y \kep_z - \kep_y \hep_z ] ,
\eqno(C17)$$

\sn where the subindex on $\N$ just emphasizes that this is now the
Poynting vector as seen in \It.  Next we use the summation
expressions (A27) and (A17) with one circular permutation of the
subindices of the form $x \ra y, \ y \ra z, \ z \ra x$ and then

$$\N \bigg| _{I_{\t}} = \hat{x} {c \over 4\pi} \Smp \Ikp \Hwp \hf 2
\hkp_x ,
\eqno(C18)$$

\sn that from Eq. (5) and as $\w'=ck'$ yields

$$\N \bigg| _{I_{\t}} = \hat{x} {\hbar c \over (2\pi)^3} \Ikk k' k_x'
= \N = c^2 \g ,
\eqno(C19)$$

\sn where the last expression serves to emphasize that we recover the
$\N$  of Eq. (C13) as was to be expected.  We need not emphasize that
Eq. (C19) vanishes when integrated over the angles. Next we proceed
to perform the inverse transformation to that of Eqs. (C4abc).  This
last takes us back from the $(k',{\bf k}')$  four-vector of \It \
into the $(k,{\bf k})$ four-vector of \Is, namely

$$\eqalignno{ k'&=k \csA-k_x \snA \cr k_x'&=k_x \csA-k\snA \cr
k_y'&=k_y \cr k_z'&=k_z &(C20)}$$

\sn with $A$ as defined in Eq. (C2). Invoking again Eq. (C5) and
after some algebra (Eq.  (C19)) yields

$$\eqalignno{
\N&=-\hat{x} \( {c \over 4 \pi} \) \( {\hbar c \over 2\pi^2} \) \Ik k
(1+\hk_x^2) \csA \snA \cr &=-\hat{x} \( {c \over 4 \pi} \) \( {8\pi
\over 3} \) \snx \int d\w \A  \cr &=\N(\t) \bigg| _{I_*} , &(C21)}$$

\sn that is the Poynting vector impinging on the object point in $S$
according to the observer at \Is. So we recover the Poynting vector
derived in (25).

\bigskip Of course both the $\N$ of Eq. (C19) and the $\N$ of Eq.
(C21) vanish at object proper time $\t=0$: that of Eq. (C19) because
of symmetry after the angle integrations and that of Eq. (C21) since
in that case the evaluation is made for vanishing proper time.  In
Eq. (C21) we have recovered the proper time dependence.  This shows
how the $(k,{\bf k})$ four-vector transformation takes us from the
$\N$  of the object point in $S$ as viewed in \Is \ that we derived
in Eq.  (25) to the corresponding $\N$ but as viewed in \It \ that
was derived in Eq. (C19).  The inverse transformation takes us back
in exactly the opposite way from the $\N$ as viewed in \It \  to the
corresponding $\N$  as viewed in \Is.

\bigskip This last example illustrates how and in what sense one can
formally at least remove and place back the time variable into the
expressions.  A more fundamental explanation follows.

\bigskip
\centerline{\bf B. Spheres of integration in $k$-space: $k$-spheres}

\bigskip Expression (26) for the momentum density $\g(\t)$ was
obtained with an integration carried out over the $k$-space of the
laboratory system \Is.  The momentum density referred however to the
estimate that an observer, at rest at $(c^2/a, 0, 0)$ of \Is, can
make concerning the momentum density the observer views should be
incident on the accelerated object.  At proper time $\t$ the object
is instantaneously at rest at the point $(c^2/a, 0,0)$ of the
inertial frame \It \ {\it that moves with respect to \Is \ with the
velocity $\bt c \hat{x}$} given in Eq. (1).

\bigskip From Eqs. (1), (2) and (C20), the last written as 

$$\eqalignno{ k'&=\gt(k-\bt k_x) \cr k_x'&=\gt (k_x-\bt k) \cr
k_y'&=k_y \cr k_z'&=k_z &(C22)}$$

\sn one can express Eq. (26) in terms of wavevectors only 

$$\g(\t)={\hbar \over (2\pi)^2} \int {d^3k \over k} k' {\bf k}' ,
\eqno(C23)$$

\sn where the primed wavevector refers to that of \It \ and again the
unprimed one is that of \Is. If subsequently the integration is
carried over $k$, the wavevector of  \Is, the result is again Eq.
(26).  However if we transform from the wavevector integration of \Is
\ to that of \It \  by means of Eq. (C5), we obtain an integral
exclusively over $k«$ of the form

$$\g(\t)={\hbar \over (2\pi)^2} \int d^3k' {\bf k}' ,
\eqno(C24)$$

\sn that when simply integrated of course vanishes because of
symmetry.  This is another way of viewing the same seemingly
paradoxical situation of Eq. (A4) versus Eq. (C12).  The
inconsistency originates in the improper integral.  When the integral
is modified by a cut-off factor a more clear picture emerges.  If we
multiply the integrand in Eq. (C23), that is equivalent to that in
Eq. (26) by the cut-off factor $\exp(-\w'/\w_c)$ and integrate over
$k$ it can readily be seen that the integral vanishes [25] .  If
after this we take the limit $\w_c \ra \infty$ the integral is left
at zero value.  However if instead we multiply by a cut-off factor of
the form  $\exp(-\w/\w_c)$  the result is 

$$\eqalignno{
\g(\t)&={\hbar \over (2\pi)^3} \int {d^3k \over k} k' {\bf k}' \exp
\( -{\w \over \w_c} \) \cr &=-\hat{x} {2 \hbar \over 3 \pi^2 c^4}
\gt^2 \bt (3 \ ! \ \w_c^4 ) , &(C25)}$$

\sn that corresponds essentially to the integral in Eq. (26) but with
the integration over $\w^3 d\w$ carried out between the integration
limits 0 and the cut-off frequency $\w_c$.

\bigskip There are clearly two different cases to consider.  (i) For
an observer at $(c^2/a, 0,0)$ of \Is, at object proper time $\t$ and
corresponding \Is \ time $t_*$, the object appears to be moving with
velocity $\hat{x}\bt c$ and with mechanical momentum $m\gt \bt c
\hat{x}$  (where $m$ is the  rest  mass) and  the  ZPF  looks
spherically symmetric with respect to the $k = 0$ point, i.e., the
origin of the $k$-space corresponding to \Is.  (ii) For an observer
that at the same object proper time $\t$ is instead located at the
point $(c^2/a, 0,0)$ of \It \  and thus instantaneously coincides and
comoves with the object, the object is of course at rest with no
momentum and no velocity and the ZPF looks spherically symmetric
around the origin of the corresponding $k'$-space that is the
$k$-space of \It.

\bigskip So, for the \Is \ observer the ZPF {\it cannot} be
spherically symmetrically distributed around the object.  However,
for the \It \ observer at object proper time $\t$ the ZPF does appear
to be spherically symmetrically distributed around the object.  We
emphasize that these two assertions, that of \Is \ and that of \It,
refer to the object at the same proper time instant $\t$.

\bigskip From the above discussion we extract the concept of the ZPF
$k$-sphere of an inertial observer which is the sphere in $k$-space
up to the cut-off radius $k_c = \w_c/c$ around the origin at $k = 0$
of the $k$-space corresponding to the observer's rest inertial
frame.  Inertial observers automatically have a unique $k$-sphere
which is the one corresponding to the inertial frame with respect to
which they are at rest.  The ZPF radiation for an inertial observer
is the ZPF radiation contained in the inertial observer's $k$-sphere
where $\w_c$ is the cut-off associated with the ZPF spectral
distribution.  If no cut-off is considered for the ZPF then the
sphere has infinite radius as we let the radius of the sphere in the
corresponding $k$-space go to infinity, namely, $k_c=\w_c/c \ra
\infty$.

\bigskip We illustrate the concept of the ZPF  $k$-sphere for the
case of the analysis performed in Section V, in particular for Eq.
(34) and related equations.  At proper time $\t=0$ and corresponding
\Is \ time $t_* = 0$, when the object instantaneously coincides with
the observer stationed in \Is \ at the point $(c^2/a, 0,0)$, this
observer claims that the ZPF  $k$-sphere is symmetrically distributed
around the object.  Therefore, when the momentum density is
calculated under this condition in general it should vanish as also
follows from Eq. (26) or Eq. (C21) when $\t=0$.  In the same way the
object momentum is viewed to be zero. This is however not the case
after a lapse of proper time $\D\t$ when the acceleration of the
hyperbolic motion has taken the object to be moving with velocity
$\beta_{\D\t} c \hat{x}$ with respect to such an observer stationed
in \Is.  For the observer in \Is \ still of course the ZPF
distribution remains the same, i.e., spherically symmetric around
himself and with a cut-off at $\w=\w_c$.  But as the object is moving
with respect to this observer, at rest at the point $(c^2/a,0,0)$ of
\Is, he would claim that the object is not located at the center of
his $k$-sphere.  So the observer at the point $(c^2/a, 0,0)$ of \Is \
at proper time $\D\t>0$, should see two things: (i) the object moving
with momentum $m_i\gamma_{\D\t}\beta_{\D\t} c \hat{x}$ and (ii) a net
non-zero momentum density of ZPF radiation $\g(\D\t)$ given by Eq.
(26), with $\t$ replaced by $\D\t$, that is impinging on the object
as argued in Sections III, IV amd V, and in Appendix A.

\bigskip In summary, every inertial observer has an associated
$k-$sphere, and he is at the central point of that sphere. Hence the
ZPF of his inertial frame is isotropically and homogeneously
distributed around him. And this is true for all inertial observers.
An entirely different question is how an inertial observer with his
associated $k-$sphere assesses the situation for a moving point from
the perspective of his inertial frame. The observer's ZPF is in
general neither homogeneously nor isotropically distributed around
the moving point because the moving point is located off-center in
the observer's $k-$sphere.

\vfill\eject
\centerline{\bf APPENDIX D:  COVARIANT APPROACH}

\bigskip The analysis of Sections IV and V and of Appendix B
considered only the momentum density $\g$ (or equivalently the
Poynting vector $\N=c^2\g$) contribution to the electromagnetic
momentum $\p$.  There is however an additional contribution to the
momentum that was neglected there.  Here we calculate that additional
contribution.  The price we pay is the need to invoke a more
abstract, not so directly intuitive, formalism.  The advantage,
besides theoretical generality, is, very importantly, a simpler
expression and a more direct interpretation for the inertial mass
$m_i$ of Eq. (30) [26].

\bigskip The standard expression for the four-momentum is

$$P^{\mu}= \( {U \over c}, \p \) .
\eqno(D1)$$

\sn In mechanics, the space part is
				 
$$\p=m\gamma {\bf v} = \gamma \p_N ,
\eqno(D2)$$

\sn where $\p_N$ is the standard newtonian momentum

$$\p_N=m{\bf v} ,
\eqno(D3)$$					

\sn and $m$ the rest mass. $U$ corresponds to the kinetic energy

$$U=\gamma m c^2 .
\eqno(D4)$$

\sn In the case of an electromagnetic field of total energy $U$ and
momentum $\p$, the four-momentum (D1)  can be expressed as an
integral over the electromagnetic energy-momentum stress tensor [27]

$$P^{\mu}= \int \Theta^{\mu \nu} d\sigma_{\nu} ,
\eqno(D5)$$

\sn where $d\sigma^{\nu}$ is a component of  a planar
three-dimensional space-like hypersurface element $d^3 \sigma$ such
that

$$d\sigma^{\nu} = \eta^{\nu} d^3 \sigma
\eqno(D6)$$

\sn and

$$d^3 \sigma = \eta^{\nu} d \sigma_{\nu}
\eqno(D7)$$

\sn with $\eta^{\nu}$ the normal unit vector to the hyperplane.  This
unit four-vector $\eta^{\nu}$ is then timelike, and in the $(-+++)$
convention we have

$$\eta^{\nu} \eta_{\nu}=-1 ,
\eqno(D8)$$

\sn and along the direction of the worldline trajectory of the object
at the point event, so
		         					          
$$\eta^{\nu} = (\gamma, \gamma \vec{\b})
\eqno(D9)$$

\sn and $\vec{\b}={\bf v}/c$ is the vectorial velocity of the object
in units of $c$.  In our case the object is viewed from the
laboratory inertial frame \Is, and at proper time $\t$ it is
instantaneously at rest in the inertial frame \It.  We thus select
$\eta^{\nu}$ as $(\gt,\gt \vec{\b}_{\t})$, with $\vec{\b}_{\t} c$ the
object velocity as measured in \Is, to be the  unit normal vector to
the hypersurface as viewed in \Is \ and then $d\sigma^{\nu}$ has both
time and space components.  In \It \ however, the object velocity at
proper time $\t$ is zero by definition and so $\eta^{\nu}=(1,0,0,0)$
and thus

$$d\sigma^{\nu}=(d^3 x',0,0,0) ,
\eqno(D10)$$

\sn where $d^3x'$ is the volume element in the three-space of \It. 
It can easily be seen, when Eq. (D7) is applied to the case of \It \
in Eq. (D10), that

$$d^3\sigma = \eta^{\nu} d\sigma_{\nu} = d^3 x'
\eqno(D11)$$

\sn is an invariant hypersurface element equal to a 3-space volume
element.  Thus $d^3\sigma$ has the same value in \Is \ and in \It. 

\bigskip The space part of the energy-momentum stress tensor
$\Theta^{\mu \nu}$ is the Maxwell  stress-tensor ${\bf T}$ with
symmetric matrix components $T_{ij} ; \ i,j = x, y, z$  of the form

$$T_{ij}={1 \over 4\pi} \[ E_iE_j + B_iB_j - \hf (E^2+B^2)
\delta_{ij} \] ,
\eqno(D12)$$

\sn and yielding for the symmetric $\Theta^{\mu \nu}$ the components

$$\eqalignno{
\Theta^{00}&={1 \over 8\pi} \( E^2 + B^2 \) = U , \cr
\Theta^{0i}&=\Theta^{i0}={1 \over 4\pi} \( \E \times \B \) _i = cg_i 
, \cr
\Theta^{ij}&=	-T^{ij}	 .							                     	    &(D13)}$$

\sn The four-momentum (D5) becomes then

$$\eqalignno{ P^0 &= \gamma \int \( {U \over c}-{\bf v} \cdot \g \)
d^3\sigma , &(D14a)  \cr
\p&=\gamma \int \( \g + {{\bf T} \cdot {\bf v} \over c^2} \)
d^3\sigma . &(D14b)}$$

\sn Expression (D14b) gives us the clue of how to correct the
noncovariant approaches of Sections IV andV and of Appendix B and
make them fully covariant.  We can do this either to the
momentum-flux approach of Section IV and V or to the momentum-content
approach of Appendix B.  The second case, namely the momentum content
approach of Appendix B is more in line with the particular
interpretation that we would like to emphasize for the inertial mass
of Eq. (30).  So we go to the momentum density $\g_*$ and the
momentum $\p_*$ of the electromagnetic radiation inside the volume of
the object as viewed from \Is \  represented in Eqs. (B7) and (B8) of
Appendix B.  The momentum of Eq. (B8) was calculated noncovariantly
and then directly from the momentum density of (B7) as equal to the
\Is \ momentum density multiplied by the volume that the object has
in \Is, namely $V_*$, mimicking the purely newtonian case, as for the
$\p_N$ of Eq. (D2) and (D3) above.  In order to obtain the proper
space part of the four-momentum (Eqs. (D1) and (D2) above) we need to
replace the momentum density expression $\g_*$ by the corrected
expression:  $\g_*+ ( {\bf T_*}\cdot {\bf v} )/c^2 $.

\bigskip When we integrate over the volume of the object in Eq.
(D14b) we use the invariant element $d^3x'$ of Eq. (D11), i.e., the
three-volume element of \It.  As the object volume was assumed so
small, the integrand was taken as constant in Eq. (B6), so just a
multiplication by the object volume $V_*=\gamma V_0$ as seen in \Is \
was required.  Here we can do the same.  Eq. (D14b) yields then
instead of Eq. (B6) the expression

$$\p_*=\gamma \( \g_*+{{\bf T_*}\cdot {\bf v_*} \over c^2} \) V_0 ,
\eqno(D15)$$

\sn where ${\bf T}$ signifies the ZPF Maxwell stress tensor (D12)
after the stochastic averaging (as in Eq. (A5)) has been performed,
i.e., more explicitly

$$T_{ij}={1 \over 4\pi} \< E_iE_j + B_iB_j - \hf (E^2+B^2)
\delta_{ij} \> ,
\eqno(D16)$$

\sn with   $i,j = x,y,z$.  The product ${\bf T}\cdot {\bf v}$ with
${\bf v} = \hat{x}v$ gives the column vector

$${\bf T}_*\cdot {\bf v}= \left[ \matrix{ T_{xx_*} v \cr T_{yx_*} v
\cr T_{zx_*} v \cr } \right] =\( \hat{x}T_{xx_*} + \hat{y}T_{yx_*} +
\hat{z} T_{zx_*} \) v .
\eqno(D17)$$

\sn It is a simple matter to show that the $y$ and $z$ components
vanish as must be expected on physical grounds.  We first show that
$T_{yx_*} = 0$.

$$T_{yx_*}={1 \over 4\pi} \< E_{y*}E_{x*}+B_{y*}B_{x*} \>
\eqno(18)$$

\sn but

$$\eqalignno{
\< E_{y*}E_{x*} \> &= \< E_{x\t} \( \gt [E_{y\t}+\bt B_{z\t} ] \) \>
\cr &=\gt \< E_{x\t}E_{y\t} \> + \gt \bt \< E_{x\t}B_{z\t} \>
&(D19)}$$

\sn and $<E_{x\t}E_{y\t}>$ involves the factor

$$\Sm \he_x \he_y = - \hk_x \hk_y
\eqno(D20)$$

\sn that vanishes upon angular integration.  In analogous fashion
$<E_{x\t}B_{z\t}>$ involves the factor

$$\Sm \he_x \ke_z = -\hk_y
\eqno(D21)$$

\sn that also vanishes when integrated over the angles.  The
derivation of $T_{zx_*}$ goes in entirely symmetric fashion and of
course also yields zero.  Next we compute $T_{xx_*}$.

$$T_{xx_*} = {1 \over 8\pi}
\< E_{x*}^2+B_{x*}^2-E_{y*}^2-B_{y*}^2-E_{z*}^2-B_{z*}^2 \> ,
\eqno(D22)$$

\sn but

$$\eqalignno{
\< E_{x*}^2 \> &= \< E_{x\t}^2 \> \cr
\< B_{x*}^2 \> &= \< B_{x\t}^2 \> \cr
\< E_{y*}^2 \> &= \< \gt \( E_{y\t} + \bt B_{z\t} \) \gt \( E_{y\t} +
\bt B_{z\t} \) \> \cr &= \gt^2 \< E_{y\t}^2 \> + \gt^2 \bt^2 \<
B_{z\t}^2 \> +2 \gt^2 \bt \< E_{y\t}B_{z\t} \> , &(D24)}$$

\sn and analogously it follows that

$$\eqalignno{ &\< B_{y*}^2 \> = \gt^2 \< B_{y\t}^2 \> + \gt^2 \bt^2
\< E_{z\t}^2 \> -2 \gt^2 \bt \< E_{z\t}B_{y\t} \>  &(D25a)  \cr &\<
E_{z*}^2 \> = \gt^2 \< E_{z\t}^2 \> + \gt^2 \bt^2 \< B_{y\t}^2 \> -2
\gt^2 \bt \< E_{z\t}B_{y\t} \>  &(D25b)  \cr &\< B_{z*}^2 \> = \gt^2
\< B_{z\t}^2 \> + \gt^2 \bt^2 \< E_{y\t}^2 \> +2 \gt^2 \bt \<
E_{y\t}B_{z\t} \> . &(D25c)} $$

\sn Furthermore

$$\< E_{i\t}^2 \> = {1 \over 3} \< E_{\t}^2 \>  = {1 \over 3} \<
B_{\t}^2 \> = \< B_{i\t}^2 \> ,
\eqno(D26)$$

\sn where $i=x,y,z$, and as

$$U={1 \over 8\pi} \< E_{\t}^2+B_{\t}^2 \> = \int \A d\w ,
\eqno(D27)$$

\sn we have that

$$\< E_{i\t}^2 \> = \< B_{i\t}^2 \> = {4 \pi \over 3} \int \A d\w
\eqno(D28)$$

\sn for $i=x,y,z$.  Hence  

$$\eqalignno{
\hat{x}{T_{xx}v \over c^2} &=
\hat{x}{1 \over c^2} c\bt {1 \over 8\pi} {4\pi \over 3} \(
2-4\gt^2-4\gt^2\bt^2 \) \int \A d\w \cr & \qquad \qquad + \hat{x}{1
\over 8\pi}{1 \over c^2} c\bt 2 \gt^2 \bt \<
E_{y\t}B_{z\t}-E_{z\t}B_{y\t} \>  \cr &=-\hat{x}{1 \over c^2} {1\over
3} c \bt \gt^2 \int \A d\w  -\hat{x}{1 \over c^2} c \bt \gt^2 \bt^2
\int \A d\w . &(D29)}$$

\sn The triangular brackets term in the second equality vanishes
because it is propotional to the ZPF Poynting vector of \It \  in the
$x$-direction and the integrations, as clearly explained in Appendix
B, should be carried over the $k$-sphere of \It.  It is furthermore
straightforward to show that each one of the summands
$<E_{y\t}B_{z\t}>$ and  $<E_{z\t}B_{y\t}>$ vanishes individually.  So

$$\g_*+{{\bf T}_* \cdot {\bf v} \over c^2}=\hat{x}{1 \over c^2} c
\bt  \int \A d\w
\eqno(D30)$$

\sn that results after a mutual cancellation of two factors of the
form $\gt^2(1-\bt^2)=1$.  We then replace Eq. (D30) into (D15) and
obtain

$$\eqalignno{
\p_*&=\hat{x} c \gt \bt {V_0 \over c^2} \int \A d\w \cr &=\hat{x} \(
{V_0 \over c^2} \int \A d\w \) c \snh . &(D31)}$$
		
\sn The inertia reaction force of Eq. (B10) becomes now

$$\eqalignno{
\Fzp_* &=-{d\p_* \over dt_*} \cr &=-{1 \over \gt}{d\p_* \over d\t }
\bigg| _{\t=0} \cr &=- \( {V_0 \over c^2} \int \ew \A d\w \)  \a \cr
&=-m_i \a , &(D32)}$$

\sn where we have written $\a = \hat{x}a$, the acceleration that goes
in the $x$-direction and have introduced again the radiation coupling
factor $\ew$.  Observe that the 4/3 factor  obtained in Sec. V and in
Appendix B becomes unity in the present case.   The inertial mass
$m_i$ is the same as Eq. (30) [28].

\bigskip The zero-component of the four-momentum we write as

$$cP^0=\gt \int (U-\g \cdot {\bf v}) d^3 \sigma \ra \gt \[ { \< E_*^2
+ B_*^2 \> \over 8\pi}-c\bt \g_* \] V_0 .
\eqno(D33)$$

\sn It is a simple matter to check that

$${ \< E_*^2 + B_*^2 \> \over 8\pi}={1 \over 3} \(
1+2\gt^2+2\gt^2\bt^2 \) \int \A d\w .
\eqno(D34)$$

\sn The $\g_*$ in Eq. (D33) is the same of (B5) that we write as

$$c\g_*={4 \over 3} \bt \gt^2 \int \A d\w .
\eqno(D35)$$

\sn From Eqs. (D33--D35) we easily obtain

$$cP^0=\gt V_0 \int \ew \A d\w = m_i c^2 \gt , 
\eqno(D36)$$

\sn which as expected is the energy of the interacting part of the
ZPF radiation inside the volume of the object $V_0$. From Eqs. (D36)
and (D31) we recover the standard mechanical four-momentum expression
for an object of rest mass $m_i$ and four-velocity

$$v^{\mu}= \( c\gt, {\bf v}\gt \) ,
\eqno(D37)$$

\sn viz,

$$P^{\mu}=m_i v^{\mu}= \( m_i c \gt, m_i {\bf v} \gt \) .
\eqno(D38)$$		

\sn This is the same four-momentum that in Eq. (33) we wrote as
$\cal{P}$.

\vfill\eject {

\bigskip

\parskip=0pt plus 2pt minus 1pt\leftskip=0.25in\parindent=-.25in 

\centerline{\bf REFERENCES}

[1] J.-P. Vigier, Foundations of Physics, {\bf 25}, No. 10, 1461 (1995).

[2] W. H. McCrea, Nature {\bf 230}, 95 (1971). See also an attempt at an alternative
approach by R. C. Jennison and A. J. Drinkwater, J. Phys. A {\bf 10}, 167 (1977);

[3] D. W. Sciama, Mon. Not. Roy. Astr. Soc. {\bf 113}, 34 (1953); see also G.
Cocconi, and E. Salpeter, Il Nuovo Cimento, {\bf 10}, (1958).

[4] S. Weinberg, {\it Gravitation and Cosmology: Principles and
Applications of the General Theory  of Relativity} (Wiley, New York,
1972), pp. 86--88.

[5] W. Rindler, Phys. Lett. A {\bf 187}, 236 (1994). W. Rindler, Phys. Lett. A
{\bf
187}, 236 (1994). There was a reply to this paper by H. Bondi and J. Samuel, Phys.
Lett. A, {\bf 228}, 121 (1997).

[6] For further detailed discussion of Mach's Principle see J.
Barbour, ``Einstein and Mach's Principle'' in {\it Studies in the
History of General Relativity}, J. Eisenstadt and A. J. Knox (eds.)
(Birkhauser, Boston, 1988), pp. 125--153.

[7] B. Haisch, A. Rueda and H. E. Puthoff, Phys. Rev. A {\bf 49}, 678
(1994).  We also refer to this paper for review points and references
on the subject of inertia.

[8] The corresponding required inertial coupling would take place
along a spacelike hypersurface in a manner consistent with the
dictates of general relativity.  See  C. W. Misner, K. S. Thorne and
T. A. Wheeler, {\it Gravitation} (Freeman \& Co., San Francisco,
1971) pp. 543--549 for a more traditional discussion on Mach's
Principle within general relativity with several references. See
however references [1], [4], [5] and [6] above.

[9] A. D. Sakharov, Sov. Phys. Dokl. {\bf 12}, 1040 (1968); Theor.
Math. Phys. {\bf 23}, 435 (1975).  See also C. W. Misner, K. S.
Thorne and  J. A. Wheeler, {\it Gravitation} (Freeman, San Francisco,
1973) pp. 417--428.  This approach has been interpreted within SED by
means of a tentative preliminary nonrelativistic treatment that
models the ZPF-induced ultrarelativistic zitterbewegung, (H. E.
Puthoff, Phys. Rev. A {\bf 39}, 2333 (1989); see also S. Carlip,
Phys. Rev. A {\bf 47}, 3452 (1993) and H.  E. Puthoff, Phys. Rev. A
{\bf 47}, 3454 (1993)).  A revision on the status of this last issue
has been carried out by K. Danley, Thesis, Cal. State Univ., Long
Beach (1994).  It shows that there remain unsettled questions in the
derivation of Newtonian gravitation.  However our inertia work
reported here and in [7] as well as the equivalence principle suggest
to us that the ZPF approach to gravitation remains promising once a
more detailed relativistic particle model is implemented.

[10] D. C. Cole and A. Rueda (1997) in preparation; and D. C. Cole
(1997) in preparation in which an effort is being made to analyze in
a more accurate way the developments of [7], in particular by not
approximating away to zero some terms, like the contribution of the
electric part of the Lorentz force, that may arguably be significant.
These involved calculations are still in progress at the time of
writing of the present paper.

[11] T. H. Boyer, Phys. Rev. D {\bf 29}, 1089 (1984); for clarity of
presentation the notation proposed in this article is followed here.

[12] W. Rindler, {\it Introduction to Special Relativity} (Oxford,
Clarendon 1991) pp. 91--93.  The most relevant part is Section 35,
pp. 90--93.  Hyperbolic motion is found in Section 14, pp. 33--36. 
Further details on hyperbolic motion are given in  F. Rohrlich, {\it
Classical Charged Particles} (Addison Wesley, Reading Mass, 1965) pp.
117 ff and 168 ff.  These are important references throughout this
paper.

[13] A. Rueda, Phys. Rev. A {\bf 23}, 2020 (1981), see e.g., Eqs. (2)
and (9).  See also D. C. Cole, Found. Phys. {\bf 20}, 225 (1990), for
a more explicit discussion of the need of a normalization factor.

[14] As the space part of the force four-vector that corresponds to
$\F+\Fzp = 0$ uniquely vanishes in all frames,  the time part should
also vanish which means that  $\v \cdot\F+\v \cdot\Fzp = 0$
in all frames (zero-component Lemma).  This together with Eq. (14)
necessarily means that the energy and momentum given by the
accelerating agent to the object is immediately passed by the object
to the surrounding vacuum field.  When the object is later
decelerated by another external agent, the energy and momentum flow
backward from the vacuum to the decelerating agent.  The proof of
this conjecture that the vacuum is the reservoir of the four-momentum
of all moving bodies requires however  critical analysis beyond the
limited scope of the present work. 

[15] See, e.g., E. J. Konopinski, {\it Electromagnetic Fields and
Relativistic Particles} (McGraw-Hill, New York 1981); or J. D.
Jackson, {\it Classical Electrodynamics} (Wiley, N.Y., 1975).

[16]	The Lorentz invariance of the spectral energy density of the
classical electromagnetic ZPF  was independently  found by T. W.
Marshall, Proc. Camb. Phil. Soc. {\bf 61}, 537 (1965) and T. H.
Boyer, Phys. Rev. {\bf 182}, 1374 (1969); see also E. Santos, Nuovo
Cimento Lett. {\bf 4}, 497 (1972). From a quantum point of view every
Lorentz-invariant theory is expected to yield a Lorentz-invariant
vacuum.  The ZPF of QED should be expected to be Lorentz-invariant,
see, e.g., T. D. Lee, ``Is the physical vacuum a medium'' in {\it A
Festschrift for Maurice Goldhaber}, G. Feinberg,  A. W. Sunyar and
J. Wenesser (eds.), Trans. N.Y. Acad. Sci., Ser. II, Vol. 40 (1980).
For nice discussions on the Lorentz invariance of the ZPF and other
comments and references to related work in SED, see L. de la Pe\~na,
``Stochastic Electrodynamics: Its development, present situation and
perspective'' in {\it Stochastic Processes Applied to Physics and Other
Related Fields} (World Scientific, Singapore, 1983) B. Gomez et al
(editors) p. 428 ff. and also L. de la Pena and A. M. Cetto {\it The
Quantum Dice} (Kluwer, Dordrecht Holland, 1996) p. 113 ff.

[17] See Appendix A of the first reference in [13].

[18] See, e.g.,  J. M. Jauch and F. Rohrlich, {\it The Theory of 
Photons and Electrons} (Springer-Verlag, Berlin, 1980), p. 298.

[19] J. Schwinger, L. L. De Rand, and K. A. Milton, Ann. Phys. (N.Y.)
{\bf 15},1 (1978), and references therein to previous work on
Schwinger's source theory.

[20] A. O. Barut, J. P. Dowling and J. F. van Huele, Phys. Rev. A
{\bf 38}, 4408 (1988);  A.O, Barut and J. P. Dowling, Phys. Rev. A 
{\bf41}, 2277 (1990); and J. P. Dowling in {\it New Frontiers in
Quantum Electrodynamics and Quantum Optics}, A.O. Barut, editor 
(Plenum, New York, 1990), and references therein to further work of
Barut and collaborators on QED based on the self-fields approach.

[21] See, however, M. Ibison and B. Haisch, Phys. Rev. A {\bf 54},
2737 (1996) for resolution of an important discrepancy between SED
and QED.

[22] J. S. Bell and J.M. Leinaas, Nucl. Phys. B {\bf 212}, 131 (1983).

[23] T. H. Boyer, Phys.  Rev. D {\bf 21}, 2137 (1980).  The time
removal procedure for SED used here is implicity found in this paper.

[24] D. C. Cole, Phys. Rev. D {\bf 35}, 562 (1987).

[25] If we insist on calculating the integral over the variables $(k,
{\bf k})$ of \Is \ instead of $(k',{\bf k}')$, i.e., those of \It ,
as long as the cut-off is spherically symmetric around the $k«$-space
origin of \It \ the end result should also be zero.  This is more
easily done by means of an exponential cut-off in a development due
to Boyer (D.C. Cole, 1993--1994 personal communication). Expression
(C23) with such cut-off reads

$$\g(\t)=\hat{x} {\hbar \over (2\pi)^2 } \int {d^3k \over k} k' {\bf
k}' \exp \( - {ck' \over \w_c} \) .$$

\sn Integration over the azimuth and over the frequency $\w=ck$
yields an expression proportional again to $\w_c^4$ and to an
integral expression on the latitude angle.  This last it can be shown
identically vanishes.  So, even if the limit $\w_c \ra \infty$ is
taken at the end, the result is still zero.

[26] All four books in references [12] and [15] may be useful
references here. The book of Rohrlich, in [12], devotes considerable
part of it to the mass problem for classical electrodynamics.  We
will need material from Rohrlich p. 86 ff and p. 129 ff.  In the book
of Jackson, in [15], this material appears in pp. 236 ff and pp. 791
ff. In Konopinski's book [15], this material is scattered through
several chapters. Rindler gives only a concise exposition of the
electromagetic energy tensor in Sec. 42 of his book [12].

[27] For $P^{\mu}$ to be a four-vector the four-divergence of the
electromagnetic energy momentum stress tensor should vanish,
$\partial_{\mu} \Theta^{\mu \nu}=0$. As the only interaction
considered in this work is the electromagnetic and as explicitly we
omit any other components of the vacuum besides the electromagnetic,
there is no question that the stress tensor $\Theta^{\mu \nu}$ is
purely electromagnetic. In more complex models where there are other
interactions it would be the four-divergence of the sum of the
electromagnetic and the other field stress tensor (Poincar\'e stress)
that should vanish, i.e., $\partial_{\mu} (\Theta^{\mu \nu} + 
\Theta^{\mu \nu}_{{\rm other}} ) = 0$. In such a case it becomes a
matter of choice if individually
$\partial_{\mu} \Theta^{\mu \nu}$ and 
$\partial_{\mu} \Theta^{\mu \nu}_{{\rm other}}$ each separately
vanishes or not, and their divergences are then just the opposites of
each other. A nice discussion of this point is found in I. Campos and
J. L. Jim\'enez, Phy. Rev. D {\bf 33}, 607 (1986) (See also I. Campos
and J. L. Jim\'enez, Eur. J. Phys. {\bf 13}, 177 (1992); T. H. Boyer,
Phys. Rev. D {\bf 25}, 3246 (1982); T. H. Boyer, Phys. Rev. D {\bf
25}, 3251 (1982)). So when there are other fields (interactions), the
electromagnetic four-vector character of $P^{\mu}$ is not that
compelling, but in the present purely electromagnetic case such
four-vector character necessarily holds since $\partial_{\mu}
\Theta^{\mu \nu}=0$. In the pure electromagnetic case that for
simplicity of treatment we assume here, the 4/3 factor becomes unity.
If, on the other hand, we were to assume other fields (e.g. those
participating in the $\ew$ response of the particle) then the
obliteration of the 4/3 factor becomes more a matter of theoretical
preference.

[28] The theory of the classical electron also presented a factor of
4/3 that could be ``corrected'' to unity by assuming a point model
for the electron with an electromagnetic energy-momentum stress
tensor of vanishing divergence. See the discussion of Ref. [27] and
the articles there for some insights on the history of the 4/3
factor; also the book of Rohrlich [12], pp. 16--18 for a detailed
scholarly account.

}

\bye